\documentclass[onecolumn,nobibnotes,nofootinbib,superscriptaddress]{revtex4}
\usepackage{amsmath,amssymb,bm}
\usepackage[a4paper,bindingoffset=0.2in,left=0.8in,right=0.8in,top=1in,bottom=1in,footskip=.25in]{geometry}
\usepackage{graphicx,subfigure,epsfig}
\usepackage{bigints}
\usepackage{color}
\usepackage[breaklinks,colorlinks,urlcolor=blue,citecolor=blue,linkcolor=blue]{hyperref}
\usepackage{mciteplus}
\definecolor{lcolor}{rgb}{0.5,0,0}
\definecolor{citcolor}{rgb}{0,0.3,0.0}
\usepackage[capitalise]{cleveref}
\usepackage{todonotes}

\newcommand{\be}{\begin{equation}}
\newcommand{\ee}{\end{equation}}
\newcommand{\beq}{\begin{eqnarray}}
\newcommand{\eeq}{\end{eqnarray}}
\newcommand{\benn}{\begin{displaymath}}
\newcommand{\eenn}{\end{displaymath}}
\newcommand{\beann}{\begin{eqnarray*}}
\newcommand{\eeann}{\end{eqnarray*}}

\begin{document}

\title{Analytic solution of Balitsky-Kovchegov equation with running coupling constant using homogeneous balance method}
\author{Yanbing Cai}
\affiliation{Guizhou Key Laboratory in Physics and Related Areas, Guizhou University of Finance and Economics, Guiyang 550025, China}
\affiliation{Institute of Modern Physics, Chinese Academy of Sciences, Lanzhou 730000, China}
\author{Xiaopeng Wang}
\affiliation{Institute of Modern Physics, Chinese Academy of Sciences, Lanzhou 730000, China}
\affiliation{School of Nuclear Science and Technology, Lanzhou University, Lanzhou 730000, China}
\affiliation{School of Nuclear Science and Technology, University of Chinese Academy of Sciences, Beijing 100049, China}
\author{Xurong Chen}
\email{xchen@impcas.ac.cn (Corresponding author)}
\affiliation{Institute of Modern Physics, Chinese Academy of Sciences, Lanzhou 730000, China}
\affiliation{School of Nuclear Science and Technology, University of Chinese Academy of Sciences, Beijing 100049, China}


\begin{abstract}
In this study, we employ the homogeneous balance method to obtain an analytical solution to the Balitsky-Kovchegov equation with running coupling constant. We utilize two distinct prescriptions of the running coupling scale, namely the saturation scale dependent running coupling and the dipole momentum dependent running coupling. By fitting the proton structure function experimental data, we determine the free parameters in the analytical solution. The resulting $\chi^{2}/\text{d.o.f}$ values are determined to be $1.07$ and $1.43$, respectively. With these definitive solutions, we are able to predict exclusive $J/\psi$ production and demonstrate that analytical solutions with running coupling are in excellent agreement with $J/\psi$ differential and total cross section. Furthermore, our numerical results indicate that the analytical solution of the BK equation with running coupling constant can provide a reliable description for both the proton structure function and exclusive vector meson production.
\end{abstract}

\maketitle

\section{Introduction}
\label{intro}
In high-energy scattering processes, the partonic density exhibits a crucial phenomenon, primarily dominated by gluonic density at the small-$x$ region. The partonic density increases rapidly with the decrease in Bjorken-$x$ owing to gluon splitting, while the overlapping gluons begin to recombine and become prevalent at high-density. This results in a balance between splitting and recombination, leading to a new state of gluon saturation \cite{Albacete:2014fwa,Blaizot:2016qgz}. The quantum chromo-dynamics effective field theory, called Color Glass Condensate (CGC) \cite{McLerran:2002wj,Iancu:2003xm,Weigert:2005us,Gelis:2010nm,kovchegov_levin_2012}, provides an efficient tool to describe measurements of observables in high-energy scattering processes, especially in the presence of saturation. The widely used dipole model offers a holistic approach to computing the observables in multiple processes, such as exclusive vector meson production \cite{Kowalski:2006hc} and proton structure function $F_2$ \cite{Albacete:2010sy}, based on gluon saturation. The CGC effective field theory provides an excellent platform to describe the phenomenon of gluon saturation and facilitates the analysis and prediction of high-energy scattering measurements.

A key ingredient in the calculation of the cross section in the dipole model is the dipole scattering amplitude, which includes all the information regarding the dipole-proton strong interactions. The dipole scattering amplitude can be obtained from phenomenological model, e.g., the GBW model \cite{Golec-Biernat:1998zce,Golec-Biernat:1999qor} and IIM model \cite{Iancu:2003ge}. These models were widely used in literature since its simplicity and ability for describing simultaneously the total inclusive and diffractive deep-inelastic scattering (DIS) cross section. For instance, the GBW model gives a good description of the early experimental data using few free parameters. Moreover, these models can naturally capture some features observed in high-energy scattering processes, such as the unitarity and geometric scaling. However, these models are inadequate in describing the wealth of high precision experimental data. Therefore, it is necessary to derive the dipole scattering amplitudes from the QCD evolution equations, which can be updated to a more elaborated version by including higher order corrections.

One of the successful evolution equations for dealing with the saturation effect is the Jalilian-Marian-Iancu-McLerran-Weigert-Leonidov-Kovner (JIMWLK) equation \cite{Jalilian-Marian:1996mkd,Jalilian-Marian:1997qno,Jalilian-Marian:1997jhx,Iancu:2001md,Ferreiro:2001qy,Iancu:2001ad,Iancu:2000hn}. The JIMWLK equation is derived by boosting the target, whose nonlinear evolution reflects the saturation in the target wavefunction. In practice, the application of the JIMWLK equation is limited due to the complexity of solving an infinite coupled hierarchies of evolution equations for the Wilson line correlators. In the mean-field approximation, the JIMWLK equation expressed with the projectile wavefunction reduces to Balitsky-Kovchegov (BK) equation \cite{Balitsky:1995ub,Kovchegov:1999yj}, an equation at leading logarithmic accuracy. The BK equation contains a nonlinear term which reduces the increase of the partonic density and ensures the dipole scattering amplitude satisfies the unitarity constraints. An important result of the BK equation is the geometric scaling of the scattering amplitude, which is verified by the experimental data of DIS in the HERA energy region \cite{Stasto:2000er}, which indirectly indicates the existence of gluon-saturated matter, thus firmly supporting the CGC theory. In recent years, the BK equation has been improved by including higher order corrections, such as the running coupling BK (rcBK) equation, which includes the contribution from quark loops \cite{Balitsky:2006wa,Kovchegov:2006vj}, the full  next-to-leading-order BK (fNLOBK) equation, which includes the contributions from quark and gluon loops, as well as the tree gluon diagrams with quadratic and cubic nonlinearities \cite{Balitsky:2007feb}. Meanwhile, the recent investigations indicate that the kinematical constraint \cite{Beuf:2014uia,Iancu:2015vea} and the target rapidity representation \cite{Ducloue:2019ezk,Xiang:2021rcy} also modify the kernel of the evolution equation, which in turn affects the speed toward saturation.

It is important to point out that all of the above-mentioned BK evolution equations are integro-differential equations with nonlinear term. It is difficult to directly solve these complex nonlinear equations. Hence, numerous efforts have been implemented to obtain solutions of the evolution equations, both analytically and numerically. In terms of the numerical solution, the BK equation and its elaborated versions have been deeply investigated \cite{Golec-Biernat:2001dqn,Albacete:2003iq,Albacete:2004gw,Albacete:2010sy,
Matas:2016bwc,Ducloue:2019jmy,Lappi:2020srm} as its simpler nature than the JIMWLK equation. While the resulting dipole scattering amplitudes from these equations give a fairly successful description of the experimental data with free parameters from the initial conditions, applications are hindered by the time-consuming nature of the evolution. So, one may hope to obtain a convenient and applicable formalism from the analytical solution. One of the early works is given by Levin and Tuchin \cite{Levin:1999mw}. They present an analytical solution for the nonlinear evolution equation with the strategy diving the equation into linear and nonlinear regions based on the critical line. Later, the authors in Ref.\cite{Iancu:2002tr} explore the geometric scaling properties of the BK equation in the saturation region. The sets of BK equations can be simplified in the saturation region, allowing us to obtain an approximate analytical solution in this region \cite{Xiang:2017fjr,Xiang:2019kre,Xiang:2021rcy}.

The nonlinear BK equation can be viewed as the combination of a linear Balitsky-Fadin-Kuraev-Lipatov (BFKL) equation with a nonlinear term which tames the growth of the gluon density. It has been shown that the nonlinear BK equation reduces to the nonlinear Fisher-Kolmogorov-Petrovsky-Piscounov (FKPP) equation when the diffusion approximation is applied to the BFKL kernel \cite{Munier:2003vc,Munier:2003sj,Munier:2004xu}. The FKPP equation is well known in statistical physics and admits the asymptotic traveling-wave solutions \cite{bramson1983convergence}. Based on this diffusion approximation, the nonlinear BK equation has been widely studied and used to describe the deep inelastic scattering \cite{kovchegov_levin_2012} (and the references therein). While the reduction of the BK equation reproduces some of the phenomenology, e.g., the geometric scaling, it is limited in its application because its solution is valid only in the vicinity of the saturation scale. To circumvent these difficulties, Marquet, Peschanski,and Soyezan proposed an iterative method to extend the solution to the non-asymptotic region \cite{Peschanski:2005ic,Marquet:2005ic}. In a recent study \cite{Levin:2022fyx}, the recurrence relations for the dipole densities are proposed. By summing the larger Pomeron loops, the scattering amplitude is found to decrease at large values of rapidity. This behavior is a consequence of the diffusion approximation to the BFKL kernel, which does not lead to the saturation both in the BK equation and in the dipole-dipole amplitude. Therefore, it is critical to develop a more realistic approximation to the BFKL kernel.

Although there are limitations in the application of the BK equation with the diffusion approximation, it is of interest to search for its solutions and investigate its properties using new approaches. In fact, the FKPP equation can be solved by using the homotopy perturbation method \cite{Saikia:2022ocs} and the homogeneous balance method \cite{Yang:2020jmt}. Applying the homogeneous balance principle, we obtain an analytical traveling-wave solution for the nonlinear BK equation with fixed coupling constant \cite{Wang:2020stj}. It is essential to acknowledge that the running of the Quantum Chromodynamics (QCD) coupling is a high-order effect. Assigning the coupling as a constant could lead to an ambiguous result. In order to obtain a comprehensive evolution equation, it is necessary to account for at least the correction arising from the running coupling constant. In this study, we employ the homogeneous balance method to the nonlinear BK equation with running coupling constant. We investigate two distinct prescriptions of running coupling scale, i.e., the saturation scale dependent running coupling and the dipole momentum dependent running coupling. By fitting the proton structure function, we obtain the definitive solution of the BK equation with running coupling constant, which can also provide a good explanation for the $J/\psi$ production. The findings suggest that the analytical solution of the BK equation with running coupling constant can effectively describe the proton structure function and the exclusive vector meson production. The current study highlights the importance of considering the correction from the running coupling constant to obtain a precise evolution equation. Furthermore, the analytical solution offers a promising avenue to predict and explain experimental findings in high-energy collisions.

\section{Analytical solution of BK equation using homogeneous balance method}

In this section we provide a detailed description of the numerical solution of the BK equation using the homogeneous balance method, which is an efficient method for constructing exact solutions of a set of nonlinear partial differential equations. In order to introduce the notation and set up the framework, we first review the general aspects of the BK equation and introduce the homogeneous balance method to obtain the analytical solution for the BK equation with fixed coupling. Then, we extend this method to the BK equation with running coupling constant.

\subsection{Analytical solution of BK equation with fixed coupling constant}

The BK equation is first derived by Balitsky \cite{Balitsky:1995ub} and Kovchegov \cite{Kovchegov:1999yj} at leading logarithmic accuracy. It describes a dipole (consist of a quark and an antiquark with transverse size $r$) evolution with rapidity $Y$ by the emission of soft gluon. In the large $N_c$ limit, it can be written as
\be
\label{eq:BK_cs}
\frac{\partial N(r,Y)}{\partial Y}  =\frac{\bar{\alpha}_s}{2 \pi}
         \int d^2z_{\bot}  \frac{r^2}{r_1^2r_2^2}
           \left [ N(r_1,Y) + N(r_2,Y) -N(r,Y)- N(r_1,Y)N(r_2,Y) \right ] \ ,
\ee
where $r_1,r_2$ is the two new daughter dipole transverse size in the evolution. Here, the coupling $\bar{\alpha}_s$ is keep fixed. As we can see, the right hand of Eq.(\ref{eq:BK_cs}) contains linear terms (the first three terms) and nonlinear term (the last term). When the size of the dipole is small enough ($N \ll 1$), the nonlinear term can be neglected, then the BK equation reduces to the linear BFKL equation \cite{Lipatov:1976zz,Mueller:1993rr}. However, in the saturation region, the nonlinear term becomes extremely significant due to the saturation effects, as it has been verified in high-energy scattering experiments. For this nonlinear equation, it is difficult to obtain an exact analytical solution. There are many efforts to obtain an approximate analytical solution in the literature, such as the semiclassical approximation \cite{Levin:1999mw,DiazSaez:2011qqj} and the traveling wave approach \cite{Beuf:2010aw}.

It should be noted that Eq.(\ref{eq:BK_cs}) is an integro-differential equation in coordinate space. The BK equation is usually translated into momentum space as it has a simple form, which is friendly for solving analytically and numerically. Through Fourier transform, Eq.(\ref{eq:BK_cs}) becomes \cite{Kovchegov:1999ua}
\be
\label{eq:BK_ms}
\frac{\partial {\cal N}(k,Y)}{\partial Y} = \bar{\alpha}_s \chi\left(-\frac{\partial}{\partial L}\right){\cal N}(k,Y)
-  \bar{\alpha}_s{\cal N}^2(k,Y)\ ,
\ee
where $L=\log{\left(k^2/\Lambda^2\right)}$. The BFKL kernel is
\be
\label{eq:BfKl_kernel}
\chi(\gamma)=2\psi(1)-\psi(\gamma)-\psi(1\!-\!\gamma)\ .
\ee

The $\partial/\partial L$ in Eq.(\ref{eq:BK_ms}) is a differential operator acting on ${\cal N}(Y,k)$. As an approximation, one can expand the BFKL kernel around the critical point $\gamma_0$. As suggested by Munier and Peschanski \cite{Munier:2003vc}, one can only keep the first three terms of the expansion to avoid the mathematical difficulties regarding the infinite-order differential equation. The truncation of the BFKL kernel to second order derivative is called diffusion approximation. A precise application of the solution in this approximation is limited as it is asymptotic. In order to extend the solution to the non-asymptotic regime, Marquet, Peschanski, and Soyezan proposed an iterative method to find the solution to the non-asymptotic region \cite{Peschanski:2005ic,Marquet:2005ic}. They found that the truncation of the BFKL kernel up to second order agrees with the numerical solution in the non-asymptotic region when the free parameters are adjusted. Consistent with Ref.\cite{Marquet:2005ic}, we assume that the diffusion approximation with free parameters can be extended to the non-asymptotic region. In the diffusion approximation, the BK equation reduces to
\be
\label{eq:BK_da}
A_0{\cal N} -{\cal N}^2 -\frac{\partial {\cal N}}{\partial \bar{Y}} -A_1 \frac{\partial {\cal N}}{\partial L}
+ A_2 \frac{\partial^2 {\cal N}}{\partial^2 L} = 0\ ,
\ee
with $\bar{Y}=\bar{\alpha}_s Y$.

The coefficients $A_p\,(p=0,1,2)$ are given by
\be
\label{eq:cofA}
A_p=\sum_{i=0}^{2-p}(-1)^i \frac{\chi^{(i+p)}(\gamma_0)}{i!\ p!} \gamma_0^i\ .
\ee

Now, Eq.(\ref{eq:BK_da}) shares the same university class as the nonlinear FKPP equation through variable transformation \cite{Munier:2003vc}. In our previous study \cite{Wang:2020stj}, we have shown that the FKPP equation can be solved by using homogeneous balance method. Homogeneous balance method is an effective approach to search for the solution of a community of the nonlinear partial equations $P\left(u, u_{x}, u_{t}, u_{x x}, u_{x t}, u_{t t}, \ldots\right)=0$ \cite{WANG1995169,WANG1996279,WANG199667}. The main idea is to choose a suitable linear combination of the heuristic solutions, which in general contains undetermined coefficients. Then make the highest nonlinear term and the highest order partial derivative term to be balanced to obtain the exact solution. A more detailed process for searching the solution of a nonlinear partial equations can be found in Ref.\cite{WANG1996279}.  Applying the homogeneous balance method, the definitive traveling-wave solution for Eq.(\ref{eq:BK_da}) is given by \cite{Wang:2020stj}
\be
\label{eq:BK_sol}
{\cal N}(L, Y) = \frac{ A_0 e^{\frac{5A_0\bar{\alpha}_s Y}{3}}}{\left\{e^{\frac{5A_0\bar{\alpha}_s Y}{6}} +e^{[-\theta+\sqrt{\frac{A_0}{6A_2}}(L-A_1 \bar{\alpha}_s Y)]} \right\}^2}\ ,
\ee
where $\theta$ is a free parameter comes from the heuristic solution in homogeneous balance method.

\subsection{Analytical solution of BK equation with running coupling constant}

In the above subsection, we have show the analytical solution for the nonlinear BK equation with fixed coupling constant. However, the phenomenological studies with the structure functions \cite{Albacete:2010sy} and particle production \cite{Chirilli:2011km,Lappi:2013zma} indicate that the BK equation with fixed coupling constant is insufficient to describe the experimental data. So, we need consider some corrections to modify the QCD evolution equation in order to get a rather good description for more and more accuracy experimental data. One of the most corrections is the running coupling correction. Using three different formulations of modified BK (kinematically constrained BK, collinearly improved BK, and target rapidity representation BK) evolution equations to numerically fit the reduced cross section data, the authors in Ref.\cite{Beuf:2020dxl} found that three different schemes give an equally good descriptions to the reduced cross section data, which means that the running coupling correction is the dominant correction among them. This founding was supported by our analytical solution in the suturation region \cite{Xiang:2022iuh}. Therefore, the exact analytical solution for the BK equation with running coupling constant is a key ingredient to understand the saturation phenomenal in high-energy scattering processes.

To include the running coupling correction, we rewrite the BK equation as
\be
\label{eq:BK_rc}
\frac{\partial {\cal N}(k,Y)}{\partial Y} = \bar{\alpha}_s (\mu) \chi\left(-\frac{\partial}{\partial L}\right){\cal N}(k,Y)
- \bar{\alpha}_s (\mu) {\cal N}^2(k,Y)\ .
\ee

Applying the diffusion approximation for Eq.(\ref{eq:BK_rc}), expanding the BFKL kernel to the second order, the BK equation with running coupling constant can be written as
\be
\label{eq:BK_rc_da}
A_0{\cal N} -{\cal N}^2 -\frac{1}{\bar{\alpha}_s (\mu)}\frac{\partial {\cal N}}{\partial Y} -A_1 \frac{\partial {\cal N}}{\partial L}
+ A_2 \frac{\partial^2 {\cal N}}{\partial^2 L} = 0\ .
\ee

Different from Eq.(\ref{eq:BK_da}) where the coupling is fixed, the coupling in Eqs.(\ref{eq:BK_rc}) and (\ref{eq:BK_rc_da}) is a scale dependent variable. At one-loop accuracy, it is given by
\be
\label{eq:afs}
\bar{\alpha}_s(\mu)=\frac{1}{b \log\frac{\mu^2}{\Lambda^2}} \ ,
\ee
where
\be
\label{eq:afs_b}
 b=\frac{11N_c-2N_f}{12N_c}\ ,
 \ee
with $N_c$ and $N_f$ is the number of color and number of flavor, respectively.

In the literature, one can choose the running coupling scale as the mother dipole size or the smallest dipole size in the evolution in coordinate space. Correspondingly, the running coupling scale is usually set as the individual dipole momentum in the momentum space. However, as an alternative approach, it is most intuitive to assume that the scale in the running coupling constant is the saturation scale, since its value implies the typical momentum of the gluon in the saturation region. In this work, we shall consider both the saturation scale prescription and the dipole momentum prescription.

In the saturation scale prescription, the running coupling becomes
\be
\label{eq:afs_qs2}
\bar{\alpha}_s(Q_s^2)=\frac{1}{b \log\frac{Q_s^2}{\Lambda^2}} \ .
\ee

The saturation scale $Q_s$ is a character of gluon saturation. It is proportional to the gluon density and increases with rapidity. Specifically, in the running coupling case, the saturation scale is a related to the rapidity through \cite{Iancu:2002tr}
\be
\label{eq:qs-Y}
\log\frac{Q_s^2}{\Lambda^2}=\sqrt{cY}\ ,
\ee
with $c$ is a free parameter.

Submitting Eqs.(\ref{eq:afs_qs2}) and (\ref{eq:qs-Y}) into Eq.(\ref{eq:BK_rc_da}), one can get
\be
\label{eq:BK_afsqs}
A_0{\cal N} -{\cal N}^2 -\frac{bc}{2}\frac{\partial {\cal N}}{\partial t} -A_1 \frac{\partial {\cal N}}{\partial L}
+ A_2 \frac{\partial^2 {\cal N}}{\partial^2 L} = 0\ .
\ee

Here we make $t=\sqrt{cY}$ for mathematical convenience to derive the analytical solution. According to the homogeneous balance method, we can get the solution of Eq.(\ref{eq:BK_afsqs}) through starting with a heuristic solution as \cite{Yang:2020jmt}
\be
\label{eq:hs_qs}
{\cal N}(L, t)=\sum_{m+n=1}^{N} a_{m+n} \frac{\partial^{(m+n)}}{\partial L^{m} \partial t^{n}} f(\varphi(L, t)) \ .
\ee

According to the homogeneous balance principle, the power of $\partial/\partial L$ and $\partial/\partial t$  in the highest order derivative term and highest order nonlinear term should be balance, respectively \cite{Yang:2020jmt,WANG1996279}. Based on this principle, we can obtain
\be
\label{eq:cof}
m=2\ , ~~~~~~~
n=0\ , ~~~~~~~\mathrm{and}~~
N=m+n=2\ .
\ee

Let the coefficient in the highest order derivative be equal to one. Then, the heuristic solution is given by
\be
\label{eq:hsqs}
{\cal N}(L,t)=f''(\varphi) \varphi_{L}^2+f'(\varphi) \varphi_{LL}+a_1 f'(\varphi) \varphi_{L} +a_0 \ .
\ee

Substituting the heuristic solution into Eq.(\ref{eq:BK_afsqs}) and using the principle that the highest power of the derivative of $\varphi$ should be balance, we get
\be
\label{eq:par_sol}
A_2f^{(4)}(\varphi)-(f''(\varphi))^2=0 \ .
\ee
Equation (\ref{eq:par_sol}) has a particular solution
\be
\label{eq:par_solf}
f=-6A_2\log \varphi \ .
\ee

Assume the formalism of $\varphi$ has the traveling wave structure
\be
\label{eq:phi_tw}
\varphi=1+e^{\alpha L+ \beta t + \theta} \ .
\ee

Substituting this ansatz solution into Eq.(\ref{eq:BK_afsqs}), collecting all terms with the same order of $f$, and setting the coefficients of each order of $f$ to be zero, we can get a set of algebra equations with the undetermined parameters. Solving these equations, we can obtain the selected results for the parameters,
\be
\label{eq:cof_qs}
a_0 =0 \ , ~~~~~
a_1 = \frac{\sqrt{A_0}}{\sqrt{6A_2}} \ , ~~~~~
\alpha = \frac{\sqrt{A_0}}{\sqrt{6A_2}} \ , ~~~~~~~\mathrm{and}~~
\beta = \frac{\frac{ \sqrt{6A_0}A_1}{\sqrt{A_2}}+5A_0}{bc}\ .
\ee

Note that $\theta$ is still a free independent parameter, which will be determined by fitting experimental data. Submitting Eqs.(\ref{eq:par_solf}) and (\ref{eq:cof_qs}) into Eq.(\ref{eq:hsqs}), the solution of the BK equation with running coupling constant in saturation scale prescription is given by
\be
\label{eq:dsol_qs}
{\cal N}(L,Y)=\frac{A_0\exp\left[2\theta + 2\frac{\left(\frac{\sqrt{6A_0}A_1}{\sqrt{A_2}}
     + 5A_0\right)\sqrt{cY}}{3bc}\right]}{ \left\{\exp\left[\theta + \frac{\left(\frac{\sqrt{6A_0}A_1}{\sqrt{A_2}}
     + 5A_0\right)\sqrt{cY}}{3bc} \right]+
      \exp\left[\frac{\sqrt{A_0}L}{\sqrt{6A_2}}\right]\right\}^2} \ .
\ee

In the dipole momentum prescription, the running coupling constant becomes
\be
\label{eq:afs_k2}
\bar{\alpha}_s(k^2)=\frac{1}{b L} \ .
\ee

Submitting  Eq.(\ref{eq:afs_k2}) into Eq.(\ref{eq:BK_rc_da}), we have

\be
\label{eq:BK_rc_k2}
A_0{\cal N} -{\cal N}^2 -bL\frac{\partial {\cal N}}{\partial Y} -A_1 \frac{\partial {\cal N}}{\partial L}
+ A_2 \frac{\partial^2 {\cal N}}{\partial^2 L} = 0\ .
\ee

This is a nonlinear partial differential equation with variable coefficient, and it is extremely difficult to solve. One can only solve it under some approximation. Following  Ref.\cite{Marquet:2005ic}, this equation can be transferred into a nonlinear ordinary differential equation through an appropriate change of variable. This can be done by assuming
\be
\label{eq:BK_rc_k2_s}
{\cal N}(L,Y)=A_0 U(s) \ ,
\ee
with a scaling variable
\be
\label{eq:BK_rc_scaling}
s = L\left(-\frac{A_{0}}{A_{1}}-\frac{1}{\tilde{c}} \sqrt{b-2 A_{1} \frac{Y}{L^{2}}}\right) \ ,
\ee
where $\tilde{c}$ is a free parameter. Inserting Eq.(\ref{eq:BK_rc_k2_s}) into Eq.(\ref{eq:BK_rc_k2}), the BK equation with running coupling becomes
\be
\label{eq:BK_ordinary}
U(s)-U^2(s)+U'(s)+\frac{A_{0}A_{2}}{A_{1}^2}U''(s)=0 \ .
\ee

Applying the same strategy as Ref.\cite{Marquet:2005ic}, the terms proportional to $1/\tilde{c}$ have been neglected.  In fact, this approximation is reasonable as the parameter $1/\tilde{c}$ is small. In our simulation in Sec.\ref{results}, we found that the value of $\tilde{c}$ is $9.89$, accordingly the value of $1/\tilde{c}$ is $0.101$. This result indicates that the approximation of discarding the terms proportional to $1/\tilde{c}$ is safe. At this approximation, the evolution equation has been slightly modified, this effect has been absorbed into the $A_{0}A_{2}/A_{1}^2$ factor. Therefore, $A_{0}A_{2}/A_{1}^2$ should be adjusted through solving Eq.(\ref{eq:BK_ordinary}). According to the homogeneous balance principle, the heuristic solution can be written as \cite{fan1998A}
\be
\label{eq:hsk}
U=a_2\omega^2 + a_1\omega+a_0 \ ,
\ee
where
\be
\label{eq:hsk_w}
\omega=\text{Tanh}(\alpha s + \theta) .
\ee

Using the same approach in the saturation scale prescription, we can get the selected results of the parameters
\be
\label{eq:cof_k}
a_0 =\frac{1}{4}\ , ~~~~~
a_1 = \frac{1}{2}\ ,~~~~~
a_2 =\frac{1}{4} \ , ~~~~~\mathrm{and}~~
\alpha = - \frac{5}{12}\ .
\ee

Noting that the $A_{0}A_{2}/A_{1}^2$ factor should be modified to $6/25$ in order to make sure that the scaling variable $s$ still satisfies Eq.(\ref{eq:BK_ordinary}). This is reasonable since the critical point may shift under the approximation and then affects the coefficient values in the expansion. Substituting the above parameter into Eq.(\ref{eq:hsk}), the approximated solution is given by
\be
\label{eq:U_sol}
U(s)=\frac{1}{4}\left[1+\text{Tanh} \left(-\frac{5s}{12}+\theta\right)\right]^2 \ .
\ee

Then, the approximate solution of the BK equation with running coupling constant in dipole momentum prescription is given by
\be
\label{eq:dsol_k}
{\cal N}(L,Y)=\frac{1}{4}A_0\left\{1+\text{Tanh} \left[\frac{5L\left(A_0 \tilde{c}+A_1 \sqrt{b-\frac{2A_1Y}{L^2}}\right)}{12A_1 \tilde{c}}+\theta\right]\right\}^2 \ .
\ee

\section{Theoretical framework of proton structure function and exclusive vector meson production}

In this work we will use our analytical solution to describe the proton structure function and particle production data. For later convenience, we review in this section the general aspects of the calculations of the proton structure function and exclusive vector meson production.

\subsection{The proton structure function}

The proton structure function at small-$x$ region is one of the valuable experimental data to test and explore the phenomenon in high-energy scattering processes. Specially, the proton structure function data are usually used as an input to determine the model parameters due to its high precision and extensive dynamical covering. In the momentum space representation, the
proton structure function can be expressed in terms of the dipole scattering amplitude
\be
\label{eq:F2}
F_{2}\left(x, Q^{2}\right)=\frac{Q^{2} R_{p}^{2} N_{c}}{4 \pi^{2}} \int_{0}^{\infty} \frac{d k}{k} \int_{0}^{1} d z\left|\Psi_\gamma\left(k^{2}, z ; Q^{2}\right)\right|^{2} T(L, Y) \ ,
\ee
where $x$ is the Bjorken-$x$, $Y=\log (1/x)$, $Q^2$ is the virtuality of the photon, and $z$ is the longitudinal momentum fraction of the incoming photon carried by quark. $\left|\Psi_\gamma\right|^{2}$ denotes the photon wave function and is given by \cite{deSantanaAmaral:2006fe}
\be
\label{eq:wave_pho}
\begin{aligned}\left|\tilde{\Psi}\left(k^{2}, z ; Q^{2}\right)\right|^{2}=&\sum_{f}\left(\frac{4 \epsilon^{2}}{k^{2}+4 \epsilon^{2}}\right)^{2} e_{f}^{2}\left\{\left[z^{2}+(1-z)^{2}\right]\left[\frac{4\left(k^{2}+\epsilon^{2}\right)}{\sqrt{k^{2}\left(k^{2}+4 \epsilon^{2}\right)}} \operatorname{arcsinh}\left(\frac{k}{2 \epsilon}\right)+\frac{k^{2}-2 \epsilon^{2}}{2 \epsilon^{2}}\right]\right. \\
&\left.+\frac{4 Q^{2} z^{2}(1-z)^{2}+m_{f}^{2}}{\epsilon^{2}}\left[\frac{k^{2}+\epsilon^{2}}{\epsilon^{2}}-\frac{4 \epsilon^{4}+2 \epsilon^{2} k^{2}+k^{4}}{\epsilon^{2} \sqrt{k^{2}\left(k^{2}+4 \epsilon^{2}\right)}} \operatorname{arcsinh}\left(\frac{k}{2 \epsilon}\right)\right]\right\}\end{aligned}\ ,
\ee
where $\epsilon^{2}=z(1-z)Q^{2}+m_f^2$, and $e_{f}$ and $m_{f}$ is the charge and mass of the quark with flavor $f$, respectively.

\subsection{Exclusive vector meson production in dipole picture}

In addition to the proton structure function, the vector meson production is also an important observable to probe the gluon saturation. In particular, the investigation of saturation is facilitated by the exclusive vector meson production processes. On the one hand, the vector meson production cross section is proportional to the square of the gluon density, which makes it more sensitive to saturation. On the other hand, the gluon distribution is easily probed by measuring the distribution of the squared momentum transfer $t$.

In the dipole model, the exclusive vector meson production in electron-proton scattering is divided into three stages. First the virtual photons from electrons split into quark-antiquark pair (the dipole). The dipole then interacts with the proton by exchanging gluons. Finally, the quark-antiquark pair recombines into a vector meson. According to factorization theory, the imaginary part of the exclusive vector meson production can be written as \cite{Marquet:2007qa}
\be
\label{im_A}
\mathcal{A}^{\gamma^* p\rightarrow Vp}_{T,L}(x,Q^2,\mathbf{q}) = \mathrm{i}\int_0^1\frac{dz}{4\pi}\int d^{2}\mathbf{r}(\Psi_{V}^{*}\Psi)_{T,L}
\mathrm{e}^{-\mathrm{i}z\mathbf{r}\cdot\mathbf{q}} T(\mathbf{r},\mathbf{q},Y)\ ,
\ee
where $\mathbf{q}$ is the momentum transfer, which is related with the squared momentum transfer through $t=-\mathbf{q}^2$. In Eq.(\ref{im_A}), the dipole amplitude was expressed as a function of the momentum transfer $\mathbf{q}$ instead of the impact parameter $\mathbf{b}$ as the experimental data are directly measured as a function of squared momentum transfer. According to MPS model \cite{Marquet:2007qa}, the momentum transfer expressed amplitude is given by
\be
\label{T_q}
T(\mathbf{r},\mathbf{q},Y) =  \sigma_0\mathrm{e}^{-B\mathbf{q}^{2}} \mathcal{N}(r,Y)\ ,
\ee
where $\sigma_0$ and $B$ are free parameters.

In this work, we shall use the analytical solution in the momentum space to predict the exclusive $J/\psi$ production. The dipole amplitude $\mathcal{N}(r,x)$ in Eq.(\ref{T_q}) can be expressed by the inverse Fourier transformation
\be
\label{N_m}
\mathcal{N}(x, \boldsymbol{r})=\frac{r^{2}}{2 \pi} \int \mathrm{d}^{2} \boldsymbol{k} \mathrm{e}^{-\mathrm{i} \mathbf{k} \cdot \boldsymbol{r}} \mathcal{N}(x, \boldsymbol{k})=r^{2} \int \mathrm{d} k k J_{0}(k \cdot r) \mathcal{N}(x, k)\ .
\ee

Another ingredient to calculate the exclusive vector meson production is the vector meson overlap wave function $(\Psi_{V}^{*}\Psi)_{T,L}$. Its transverse and longitudinal components can be written as \cite{Kowalski:2006hc}
\be
\label{eq:overlapT}
  (\Psi_V^*\Psi)_{T} = \hat{e}_f e \frac{N_c}{\pi z(1-z)}\left\{m_f^2 K_0(\epsilon r)\phi_T(r,z)
- [z^2+(1-z)^2]\epsilon K_1(\epsilon r) \partial_r \phi_T(r,z)\right\}\ ,
\ee
\be
\label{eq:overlapL}
  (\Psi_V^*\Psi)_{L} = \hat{e}_f e \frac{N_c}{\pi}2Qz(1-z)K_0(\epsilon r)
\Big[M_V\phi_L(r,z)+ \delta\frac{m_f^2 - \nabla_r^2}{M_Vz(1-z)}
    \phi_L(r,z)\Big]\ ,
\ee
where $e=\sqrt{4\pi \alpha_{em}}$ and  $M_V$ is the mass of vector meson. Here $\phi(r,z)$ is the scalar function. We shall use the boosted Gaussian scalar functions since it give a rather good description for exclusive $J/\psi$ production.
\be
\label{eq:BGT}
\phi_{T}(r,z) = \mathcal{N}_{T} z(1-z)\exp\left[-\frac{m_f^2 \mathcal{R}_{T}^2}{8z(1-z)}
- \frac{2z(1-z)r^2}{\mathcal{R}_{T}^2} + \frac{m_f^2\mathcal{R}_{T}^2}{2}\right] \ ,
\ee
\be
\label{eq:BGL}
\phi_{L}(r,z) = \mathcal{N}_{L} z(1-z)\exp\left[-\frac{m_f^2 \mathcal{R}_{L}^2}{8z(1-z)}
- \frac{2z(1-z)r^2}{\mathcal{R}_{L}^2} + \frac{m_f^2\mathcal{R}_{L}^2}{2}\right]\ .
\ee
For  $J/\psi$, $N_{T}=0.578$, $N_{T}=0.575$, and $R_{T,L}=2.3$ \cite{Kowalski:2006hc}.

Taking into account the real part contribution and the skewness effect correction, the differential cross section for the exclusive vector meson production is given by
\be
\label{dif_cross}
\frac{d\sigma^{\gamma^* p\rightarrow Vp}_{T,L}}{dt}
=\frac{(1+\beta^{2})R_{g}^{2}}{16\pi}\mid\mathcal{A}^{\gamma^* p\rightarrow Vp}_{T,L}(x,Q^2,\mathbf{q})\mid^{2},
\ee
where $\beta^{2}$ factor denotes the contribution from the real part of the scatting amplitude, and the skewness effect contribution is denoted by the $R_{g}$ factor. These two factors can be expressed in terms of the imaginary part
\be
\label{corrections}
\beta = \tan\left(\frac{\pi\xi}{2}\right),~~\mathrm{and}~~
R_g= \frac{2^{2\delta+3}}{\sqrt{\pi}}\frac{\Gamma(\xi+5/2)}{\Gamma(\xi+4)}\ ,
\ee
with
\be
\label{delta}
\xi = \frac{\partial\ln(\mathcal{A}_{T,L}^{\gamma^* p\rightarrow Vp})}{\partial\ln(1/x)}\ .
\ee

The total cross section can be written as
\be
\label{total_cross}
\sigma^{\gamma^* p\rightarrow Vp}_{T,L}
=\frac{(1+\beta^{2})R_{g}^{2}}{16\pi B_D}\mid\mathcal{A}^{\gamma^* p\rightarrow Vp}_{T,L}(x,Q^2,\mathbf{q})\mid_{t=0}^{2} \ ,
\ee
where $B_D$ is the diffraction slope parameter \cite{Cepila:2019skb}.


\section{Numerical results}
\label{results}

Our fitting dataset is the proton structure function from H1 and ZUES Collaborations \cite{H1:2009pze,H1:2013ktq}. As the BK equation is a theory framework for describing the small-$x$ ($x\leq 0.01$) behavior, the data with $x> 0.01$ are excluded. Moreover, we only consider the data with $1 <Q^2 <45\,\text{GeV}^2$. The lower cut on $Q^2$ is selected to ensure that it is in the perturbative region. The upper cut on $Q^2$ is to prevent overly high values which should include the corrections from Dokshitzer-Gribov-Lipatov-Altarelli-Parisi (DGLAP) evolution.

The definitive solution for the BK equation with running coupling constant is obtained by fitting the proton structure function $F_2$. To demonstrate the influence of the choice of two different prescriptions for running coupling scale, we fix $R_p=3.2 \,\text{GeV}^{-1}$, $\Lambda=0.2 \,\text{GeV}$, and only fit the parameters from the dipole scattering amplitude in Eqs.(\ref{eq:dsol_qs}) and (\ref{eq:dsol_k}). In the fitting, we have $85$ experimental data. The parameters and $\chi^{2}/\text{d.o.f}$ results are shown in Table ~\ref{table:1}. From the $\chi^{2}/\text{d.o.f}$, we can see that the saturation scale prescription (SSP) gives a slightly better description for the structure function than the dipole momentum prescription (DMP).

\begin{table}[h!]
  \begin{center}
  \caption{Parameters and $\chi^{2}/\text{d.o.f}$ results from the fit to $85$ proton structure function experimental data.}
  \begin{tabular}{ccccccccc}
  \hline
  &      & $A_0$ & $A_1$ & $A_2$ & $c$ & $\tilde{c}$ & $\theta$ & $\chi^{2}/\text{d.o.f}$ \\
  \hline
  & SSP    & 6.98   & -8.46   & 2.95  & 3.11   & $-$   & -2.76  & 1.07  \\

  & DMP    & 22.20   & -36.06   & 14.06  & $-$   & 9.89   & -1.66  & 1.43 \\
  \hline
    \end{tabular}%
  \label{table:1}
  \end{center}
\end{table}

\begin{figure}[h!]
\begin{center}
\includegraphics[width=0.31\textwidth]{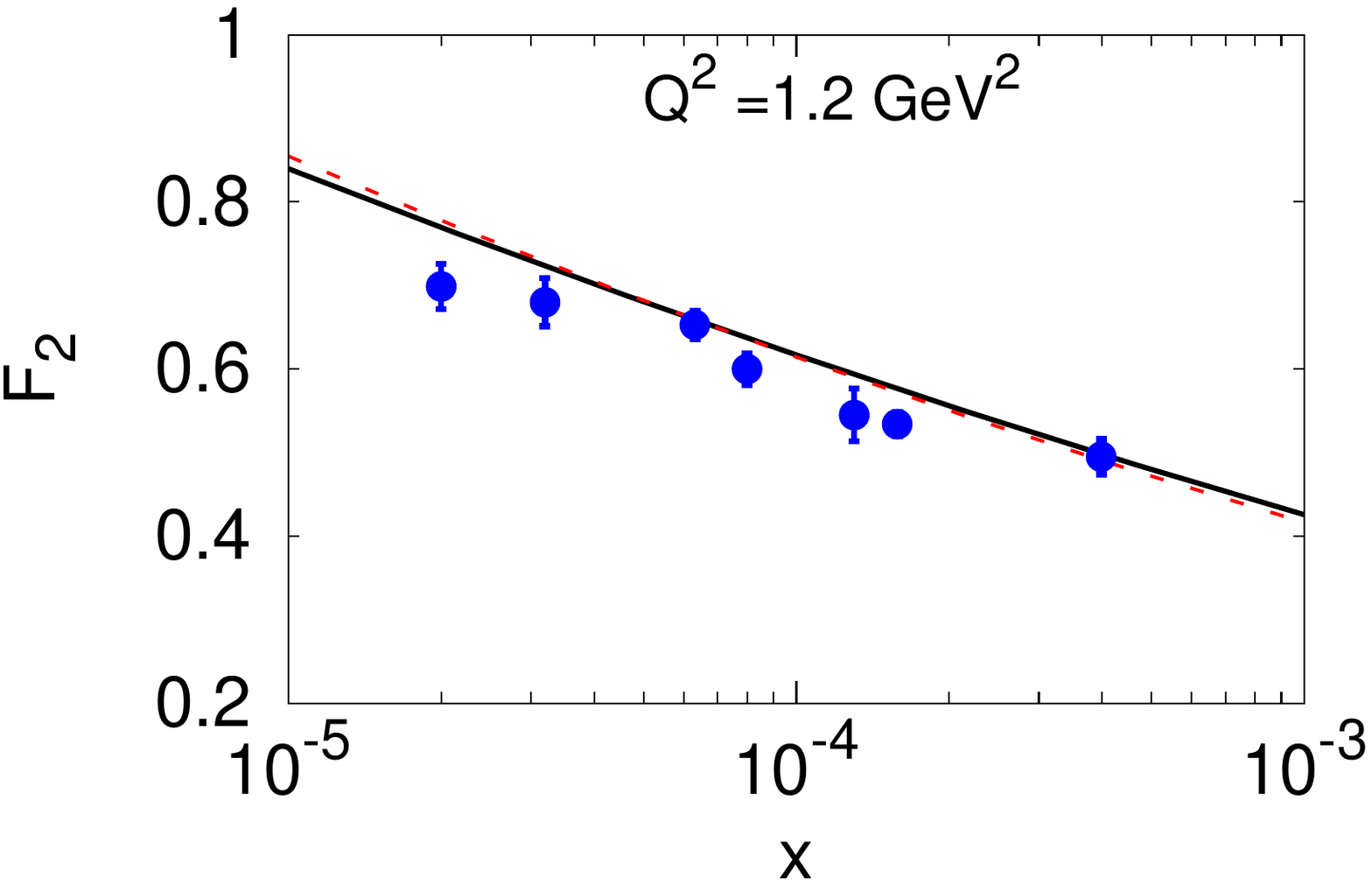}
\hspace{-0.3cm}
\includegraphics[width=0.31\textwidth]{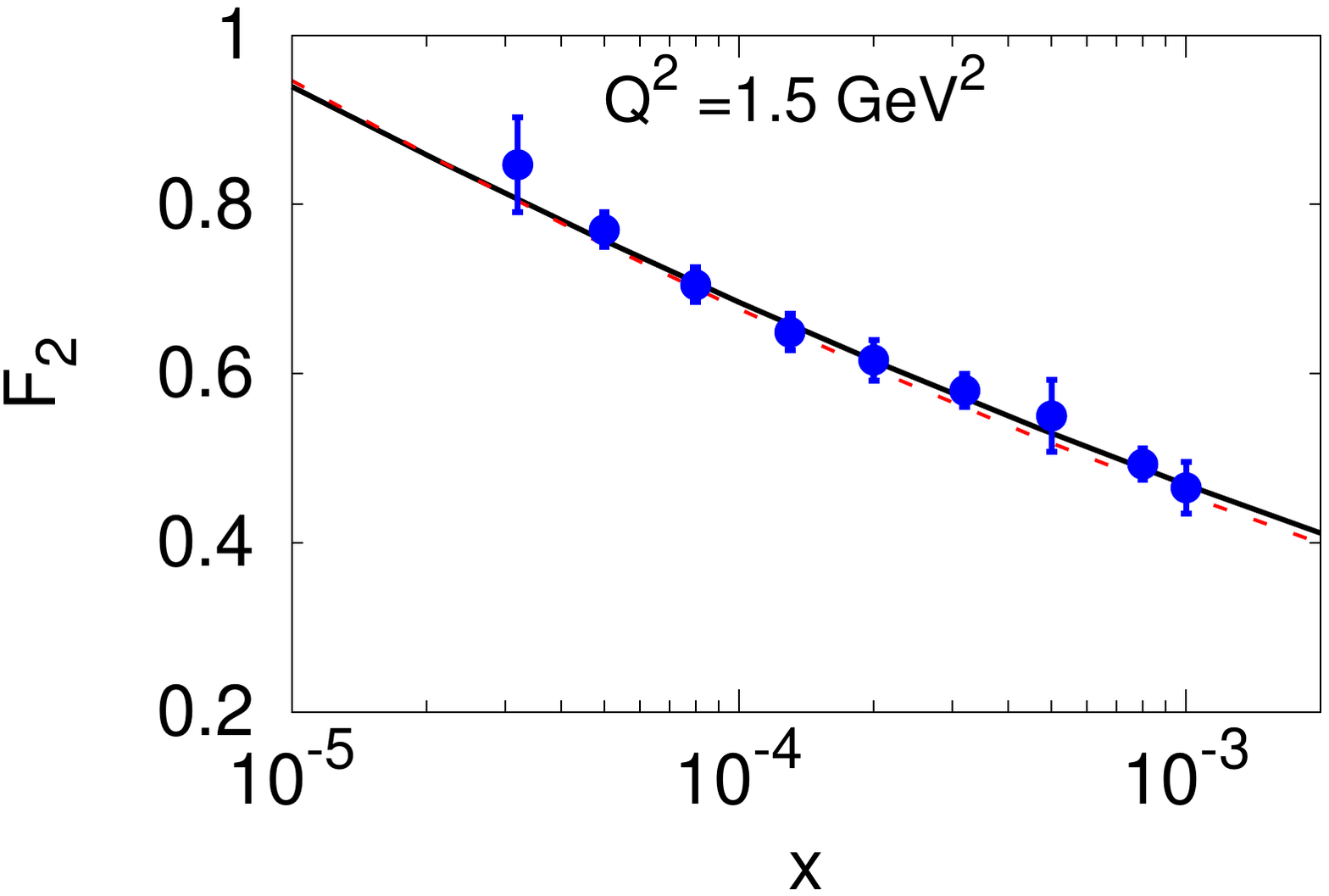}
\hspace{-0.3cm}
\includegraphics[width=0.31\textwidth]{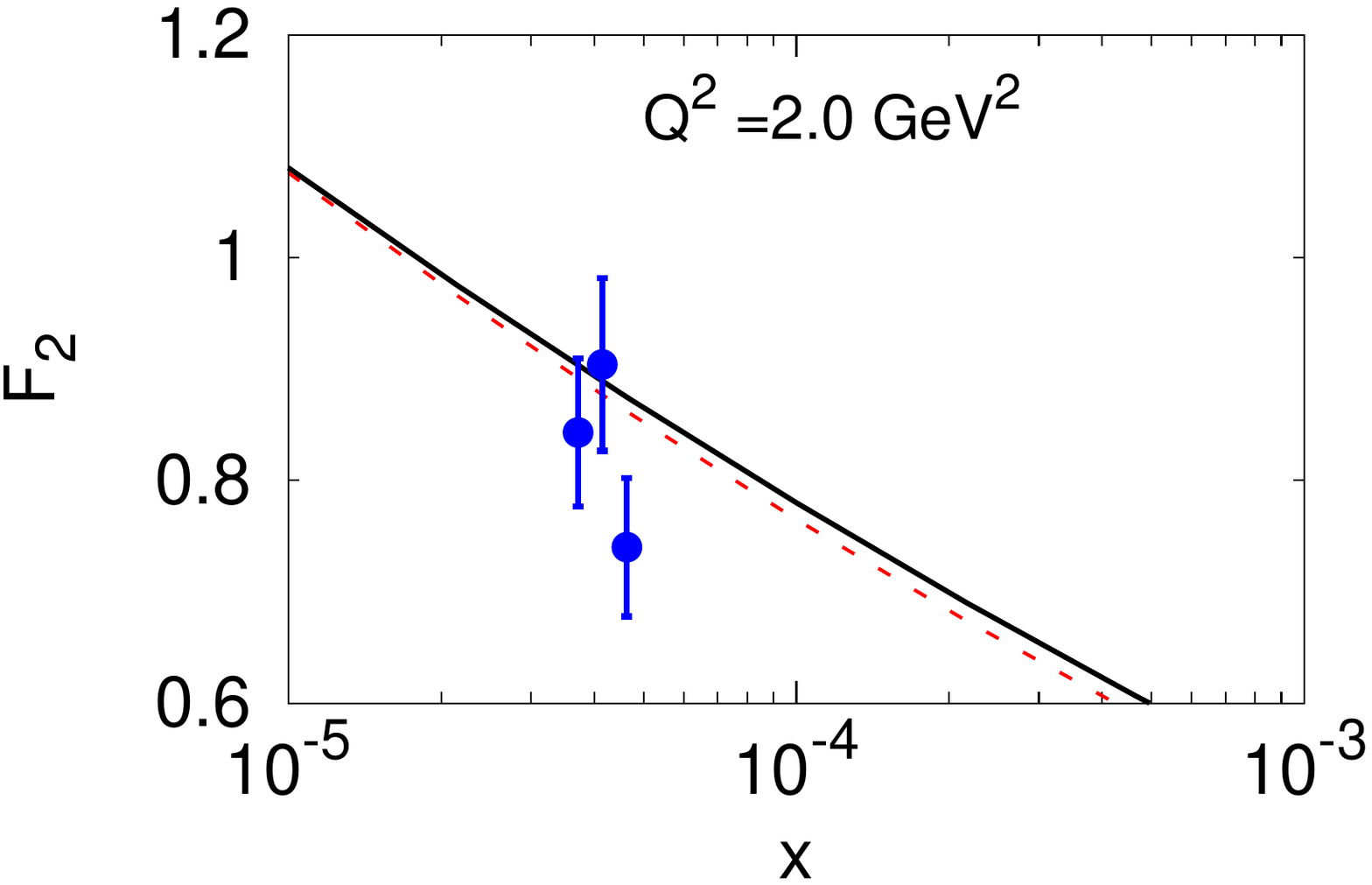}
\includegraphics[width=0.31\textwidth]{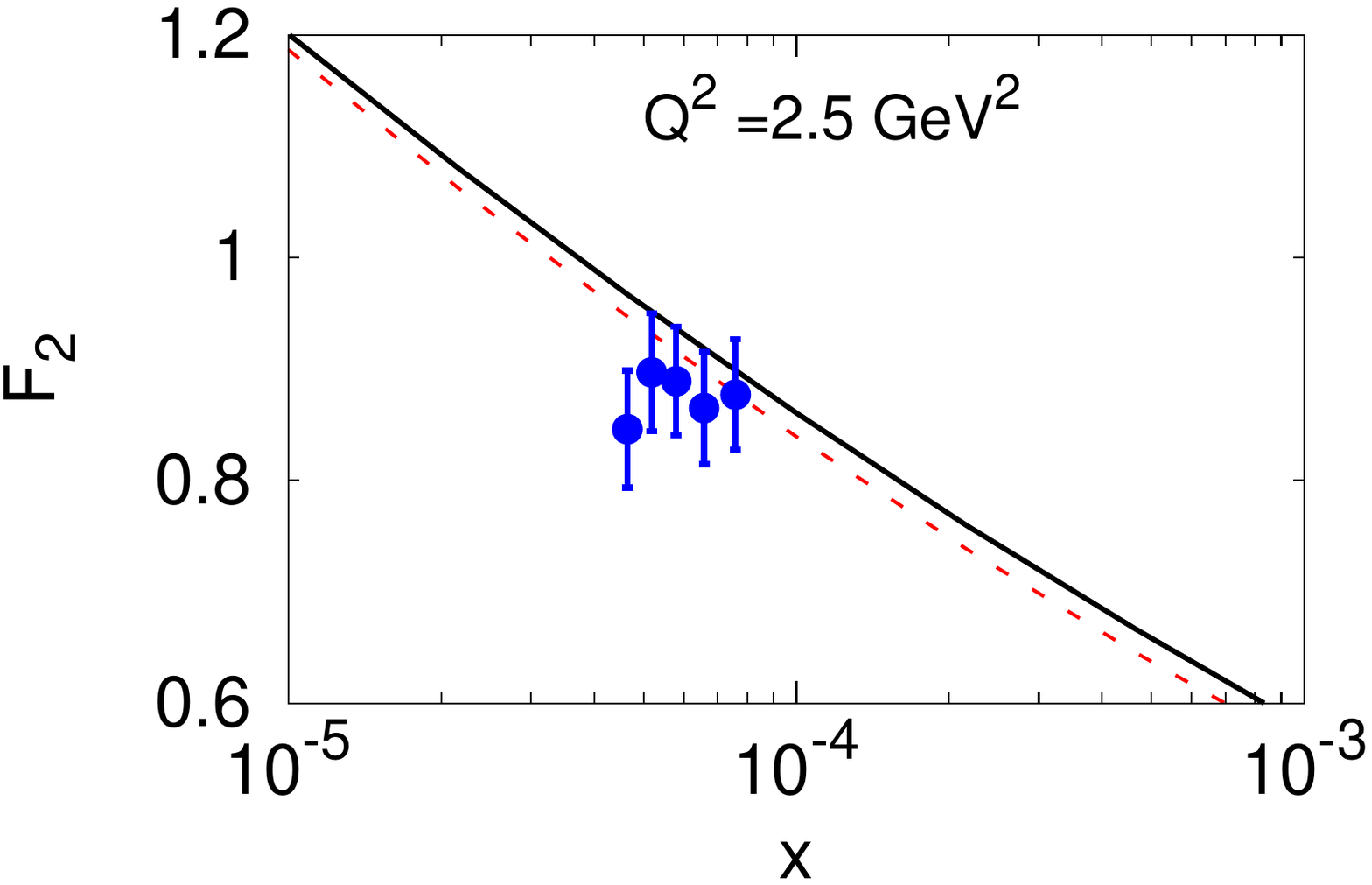}
\hspace{-0.3cm}
\includegraphics[width=0.31\textwidth]{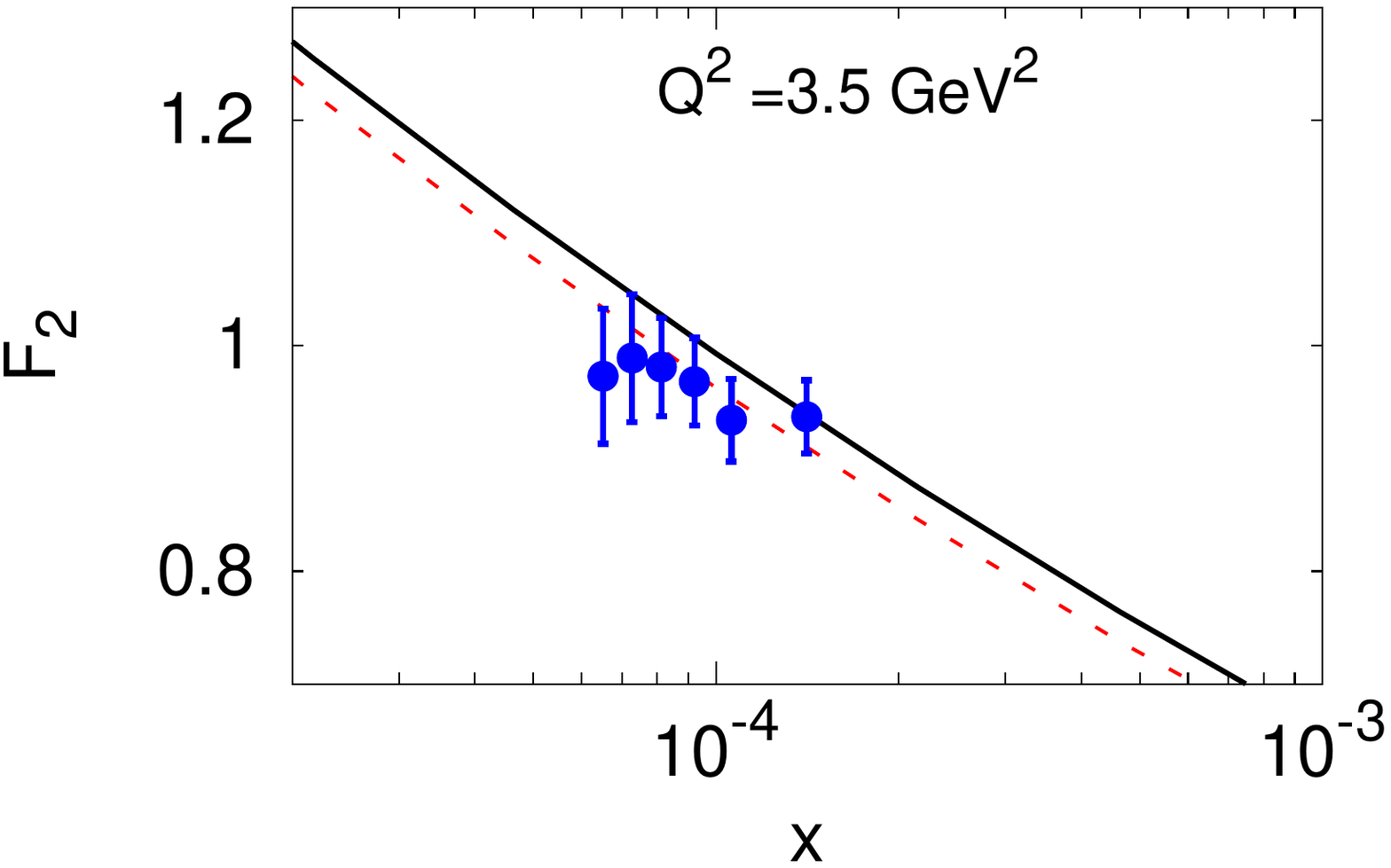}
\hspace{-0.3cm}
\includegraphics[width=0.31\textwidth]{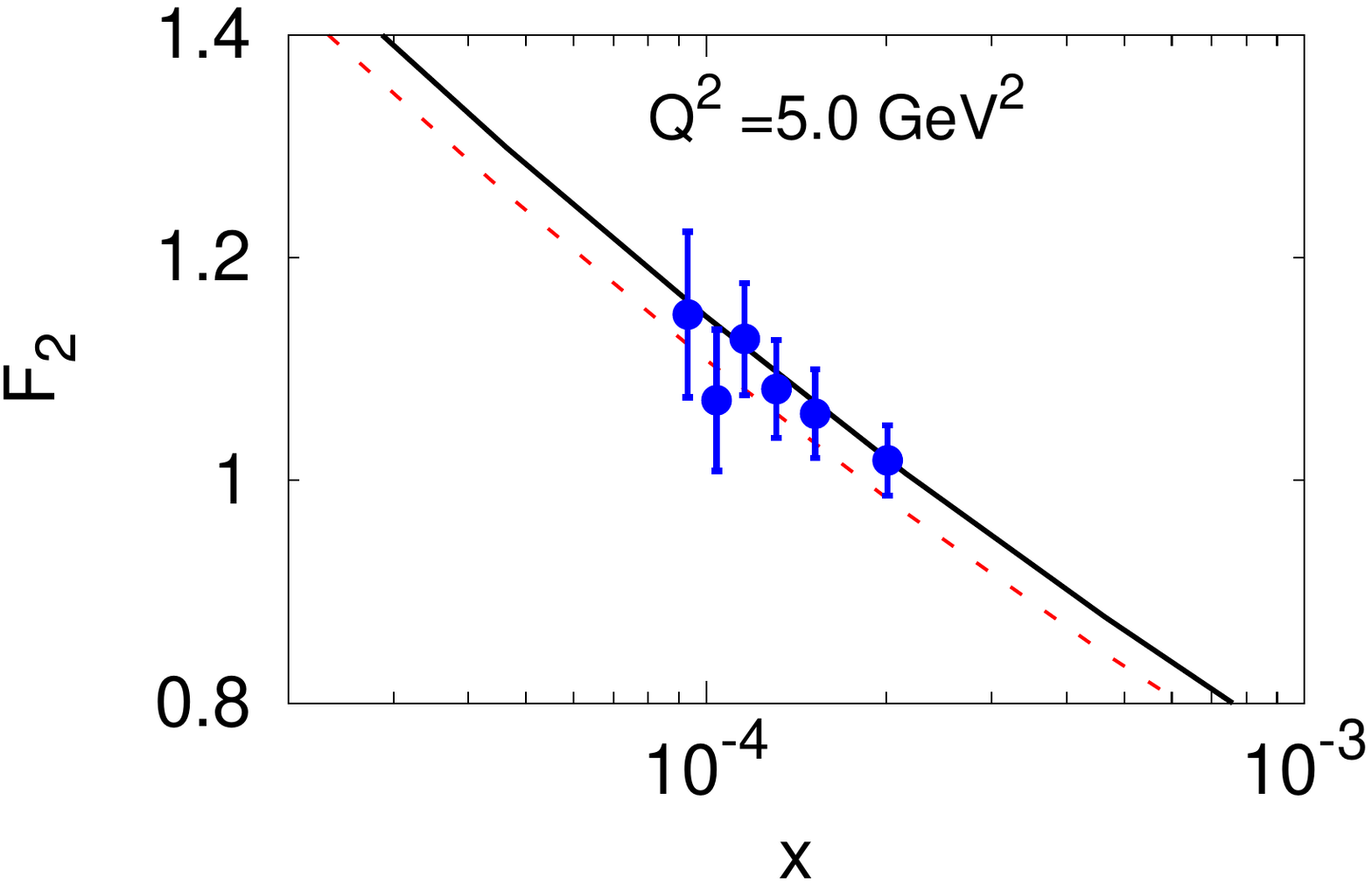}
\includegraphics[width=0.31\textwidth]{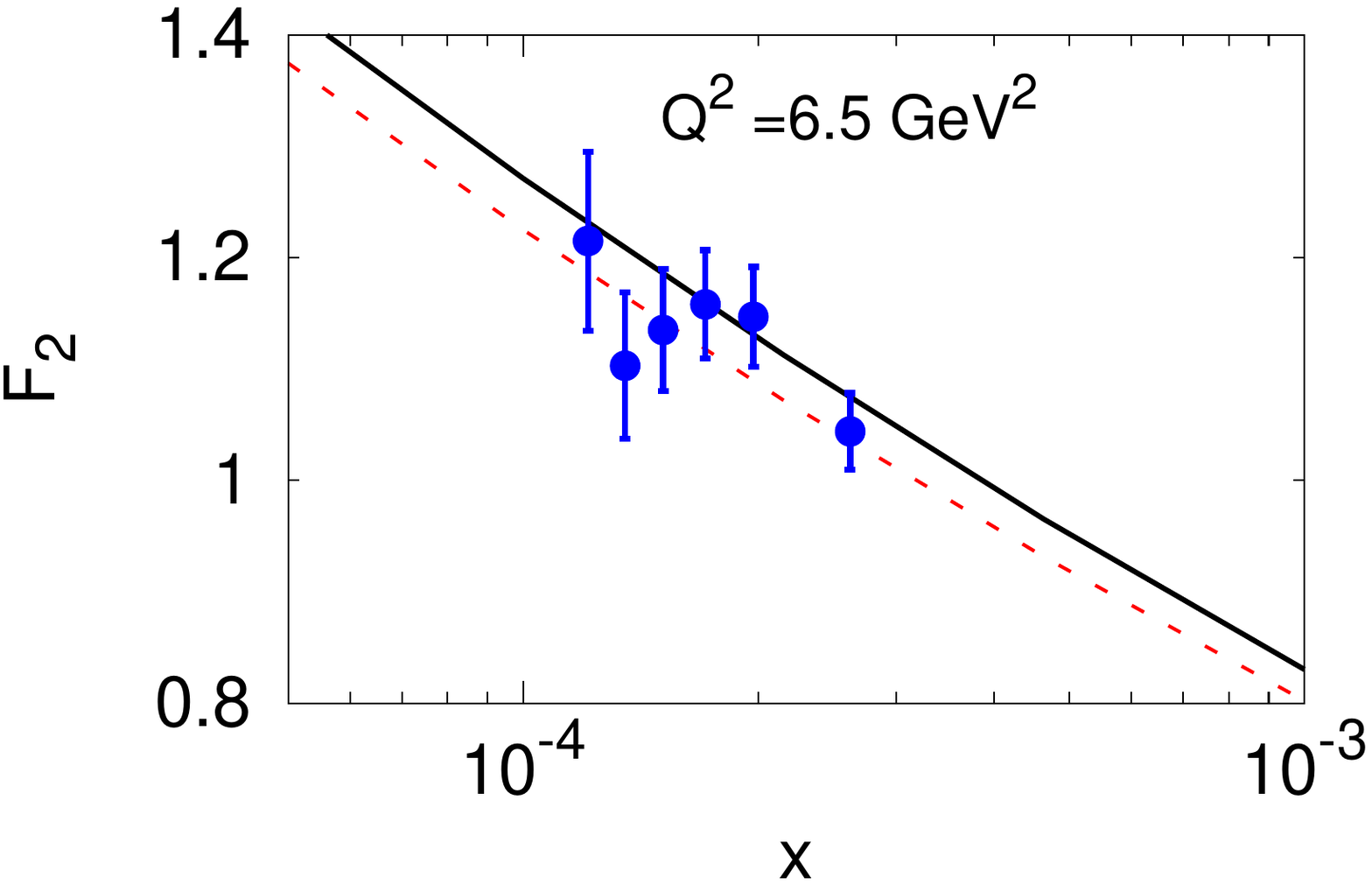}
\hspace{-0.3cm}
\includegraphics[width=0.31\textwidth]{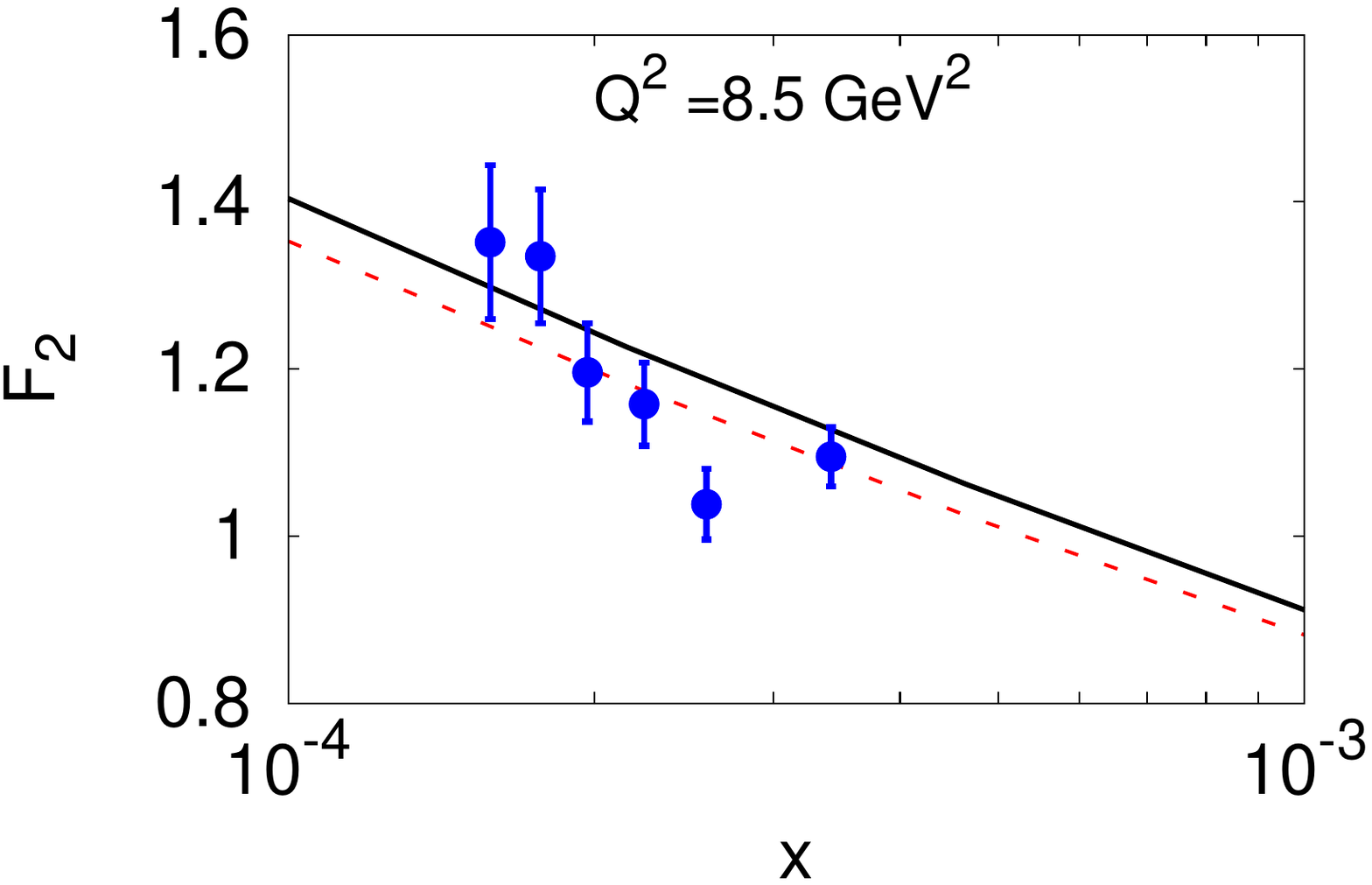}
\hspace{-0.3cm}
\includegraphics[width=0.31\textwidth]{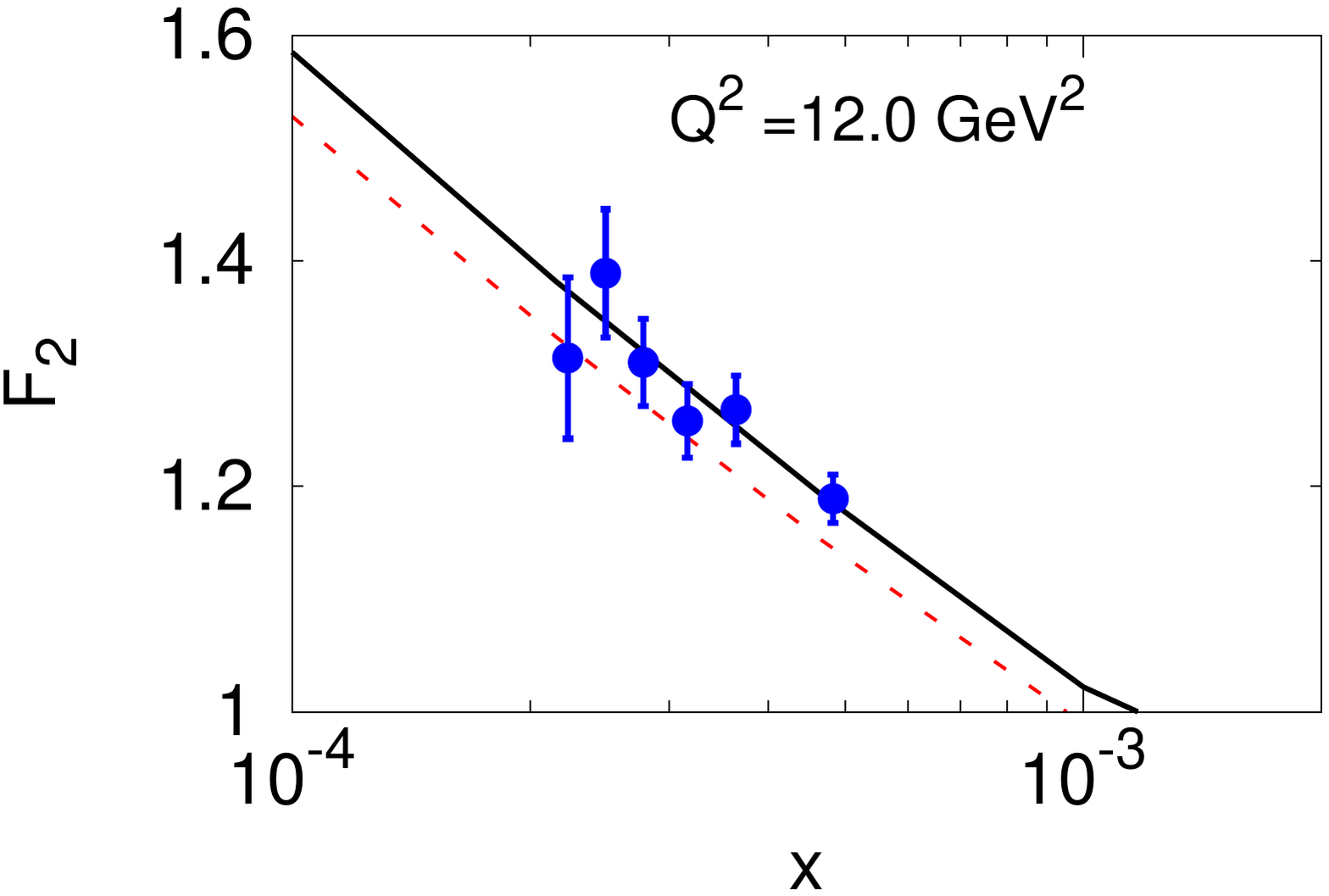}
\includegraphics[width=0.31\textwidth]{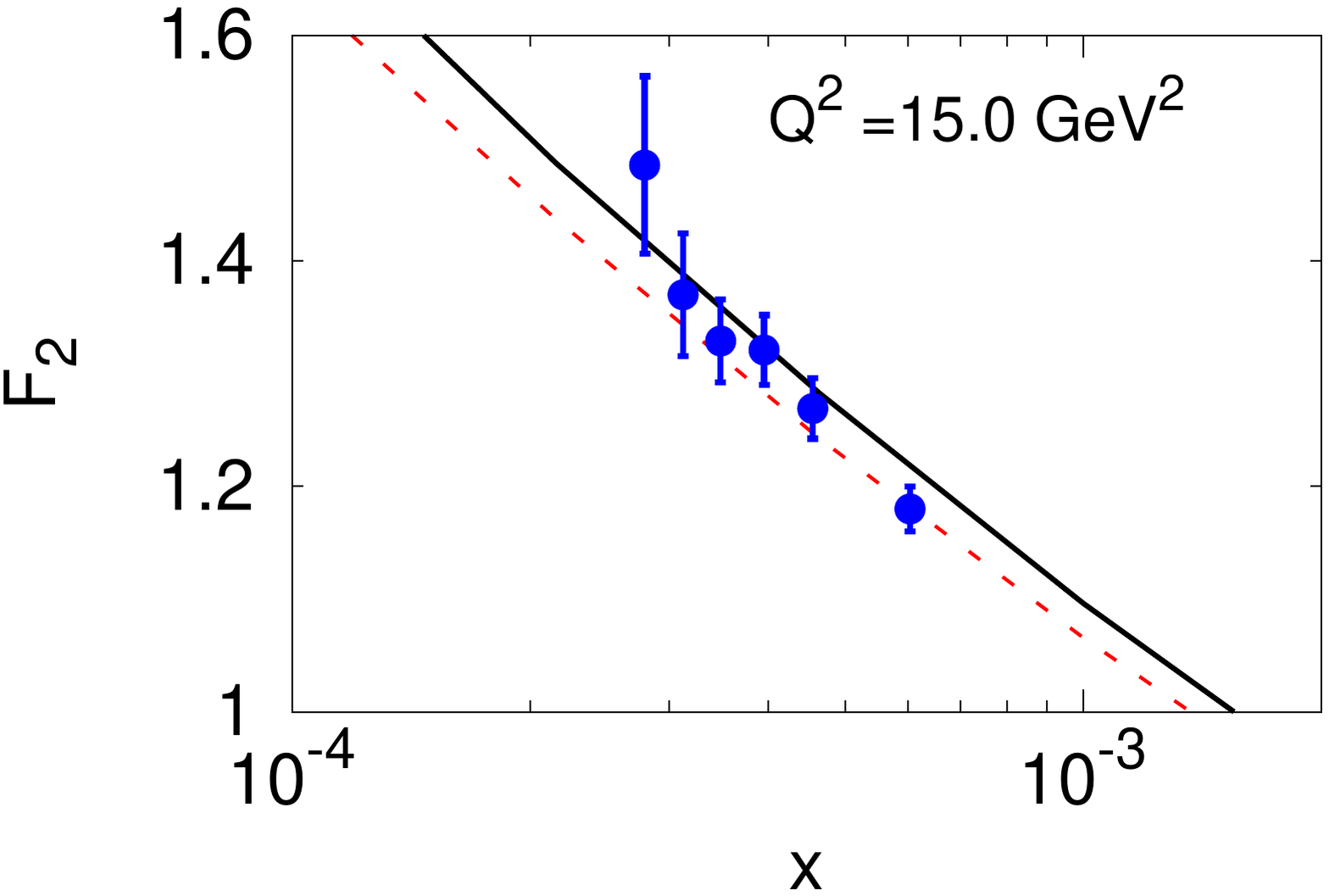}
\hspace{-0.3cm}
\includegraphics[width=0.31\textwidth]{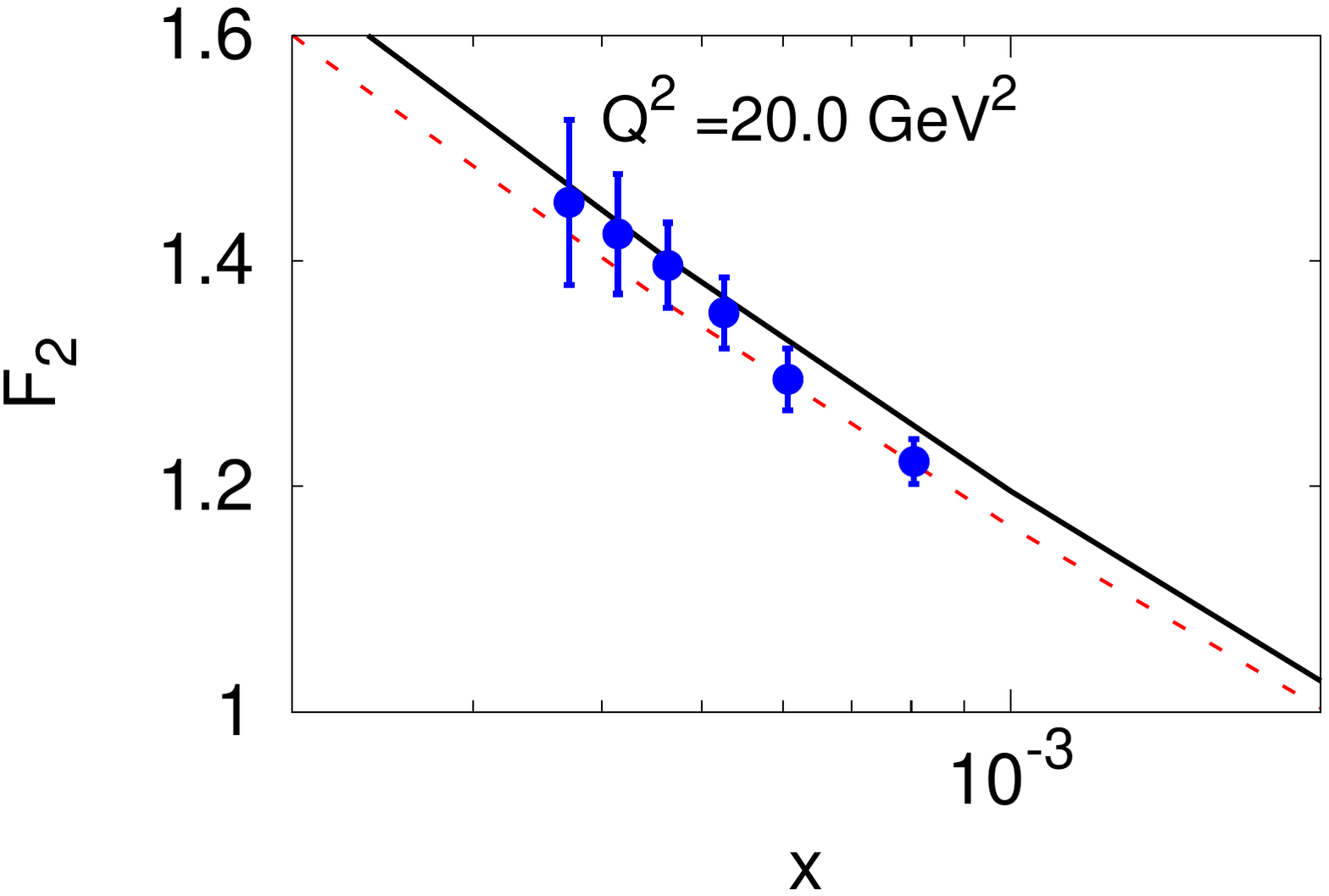}
\hspace{-0.3cm}
\includegraphics[width=0.31\textwidth]{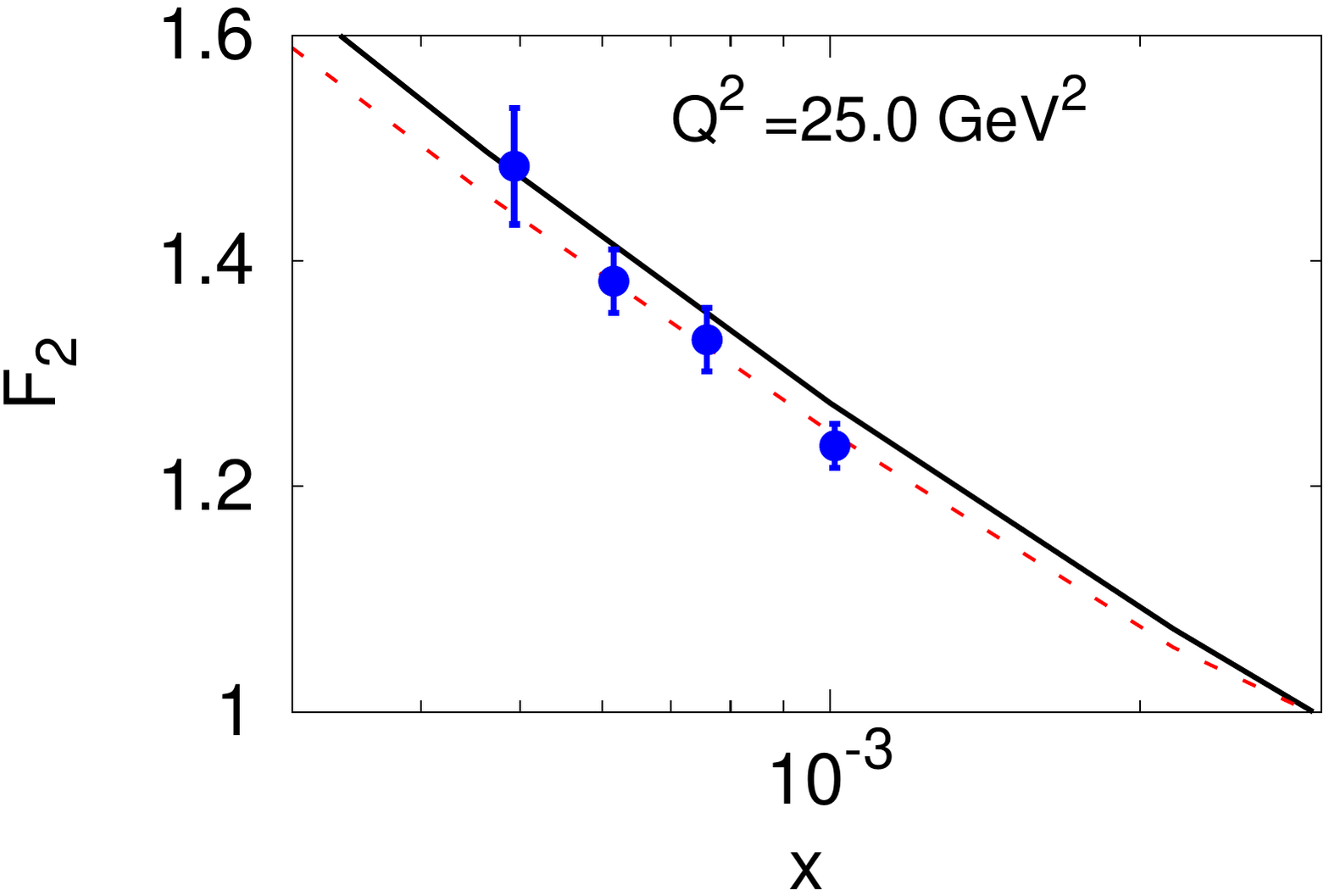}
\includegraphics[width=0.31\textwidth]{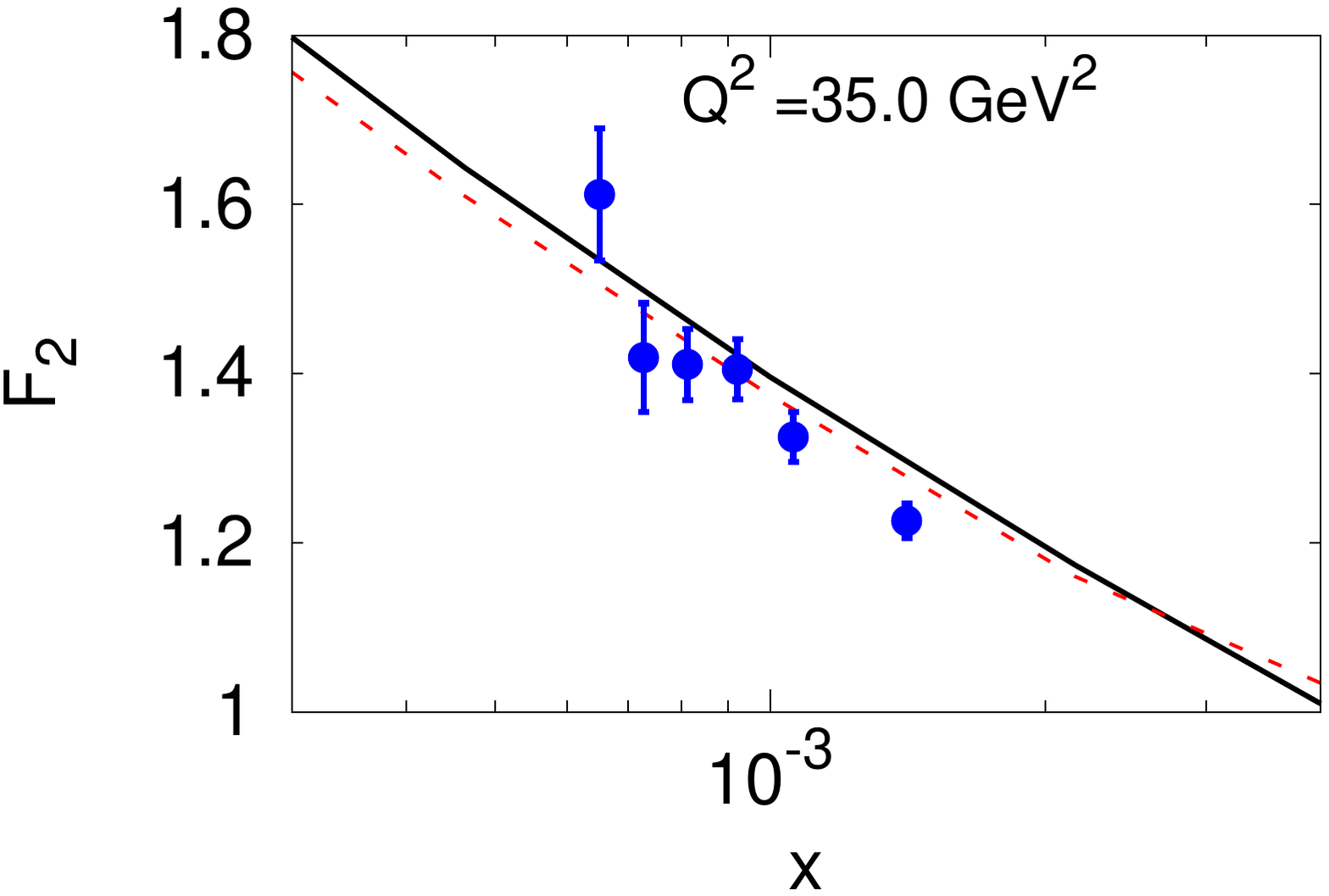}
\hspace{-0.3cm}
\includegraphics[width=0.31\textwidth]{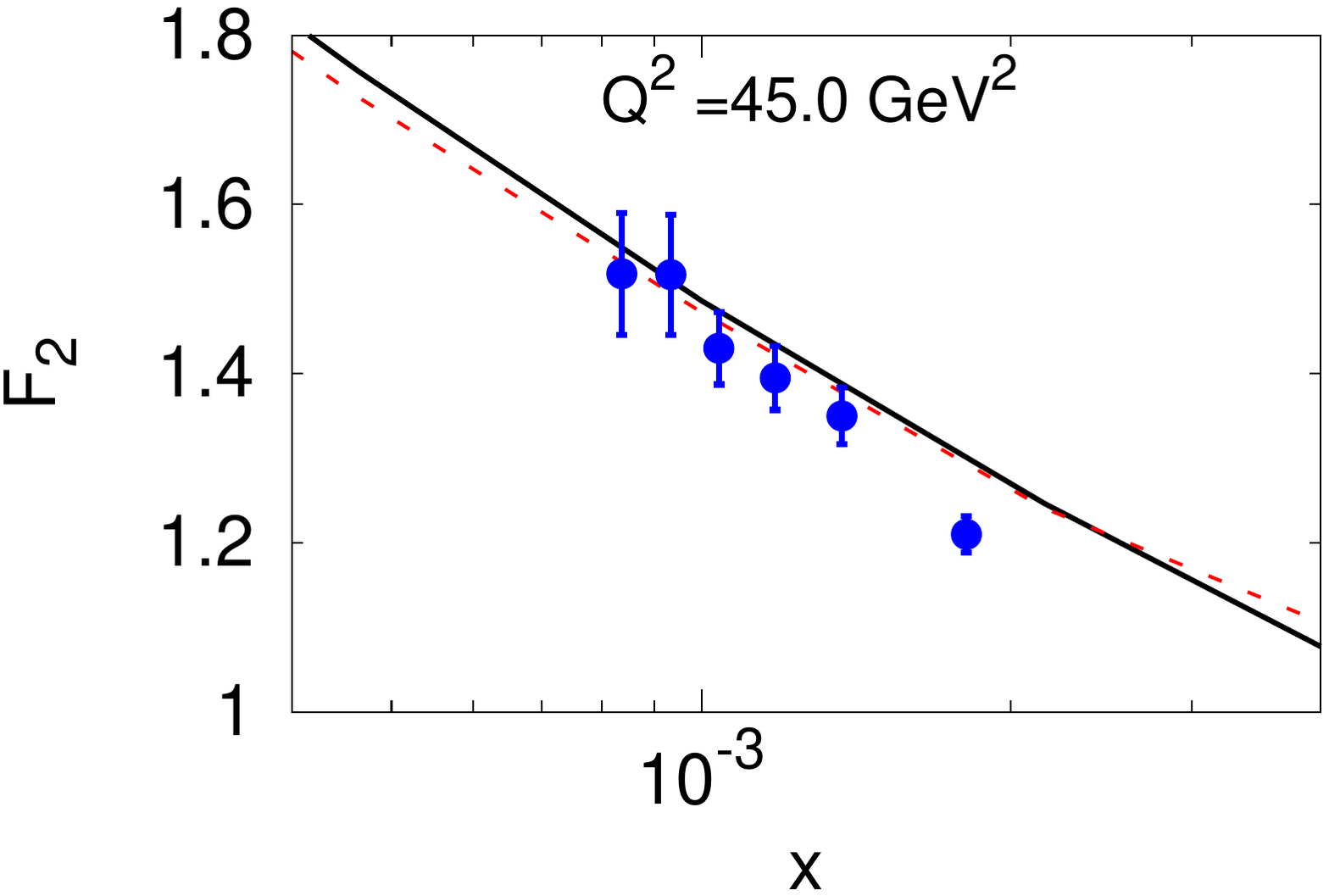}
\hspace{-0.3cm}
\includegraphics[width=0.31\textwidth]{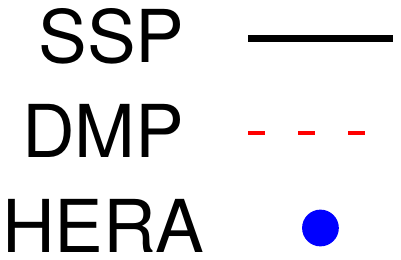}
\end{center}
\vspace{-0.5cm}
\caption{The proton structure function $F_2$ as a function of Bjorken-$x$ at different $Q^{2}$. The solid lines and the dashed lines are fitting from saturation scale prescription (SSP) and the dipole momentum prescription (DMP), respectively. The data are from  H1 and ZUES Collaborations \cite{H1:2009pze,H1:2013ktq} at HERA. }
\label{fig:F2}
\end{figure}

Figure \ref{fig:F2} shows the fitting results of proton structure function as a function of Bjorken-$x$ at different virtuality $Q^{2}$. The solid lines and the dashed lines are fitting from saturation scale prescription and the dipole momentum prescription, respectively (similarly hereinafter). The data are from  H1 and ZUES Collaborations \cite{H1:2009pze,H1:2013ktq} at HERA. As we can see, both prescriptions give similar trends for the proton structure function. However, the saturation scale prescription provides a rather better description in high $Q^{2}$. It means that the saturation scale prescription is more favored by the experimental data, which is consistent with the $\chi^{2}/\text{d.o.f}$ in Table ~\ref{table:1}. The slightly better performance for the saturation scaling prescription can be attributed to the two facts that saturation scale is the typical momentum of the gluon in the saturation and Eq.(\ref{eq:dsol_qs}) is obtained without too great approximation.

\begin{figure}[t!]
\begin{center}
\includegraphics[width=0.5\textwidth]{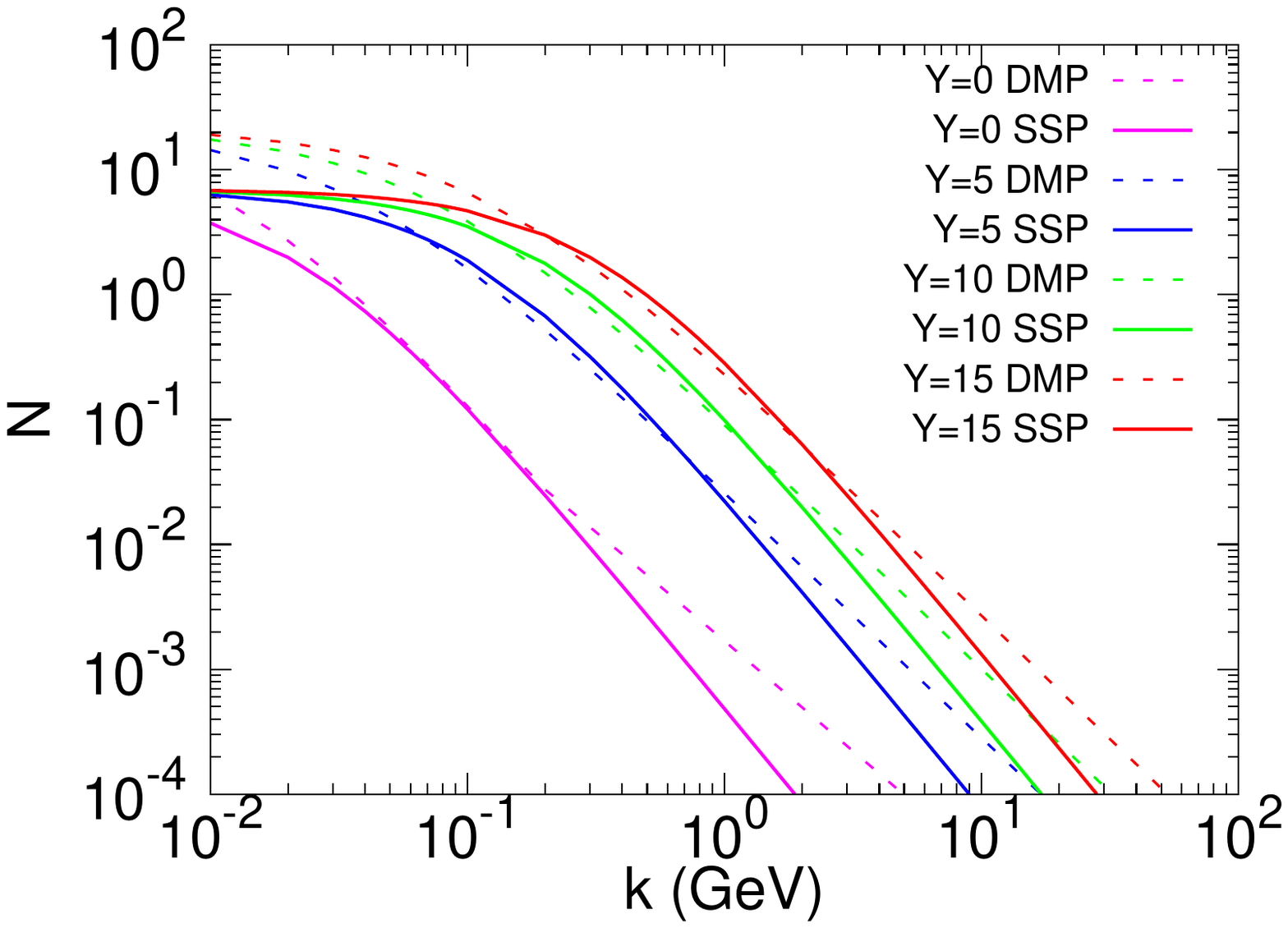}
\vspace{-0.5cm}
\end{center}
\caption{The dipole scattering amplitudes from the solution of the BK equation with running coupling constant in saturation scale prescription (SSP) and dipole momentum prescription (DMP), respectively.}
\label{fig:dip}
\end{figure}

Figure \ref{fig:dip} presents the dipole scattering amplitudes of Eqs.(\ref{eq:dsol_qs}) and (\ref{eq:dsol_k}) with parameters from Table ~\ref{table:1}. It is clear that the dipole scattering amplitude is scale dependent in the running coupling case. However, the values of dipole scattering amplitude have little difference in certain regions as is shown in Fig.\ref{fig:dip}. Therefore, both prescriptions can give a good description for the proton structure function.

Exclusive vector meson production is sensitive to saturation physics and can be an excellent probe of gluon saturation. The study of its distribution has attracted a great deal of interest. As it was shown that the exclusive vector meson production can be well described in the dipole model \cite{Kowalski:2006hc}. To test the robustness of our analytic solution, we use the solutions from Eqs.(\ref{eq:dsol_qs}) and (\ref{eq:dsol_k}) with the same parameters fitting from the proton structure function to predict the exclusive $J/\psi$ production in the framework of dipole model. For comparison, we also present the results obtained from the numerical solutions of the running coupling BK equation.

\begin{figure}[h!]
\begin{center}
\includegraphics[width=0.48\textwidth]{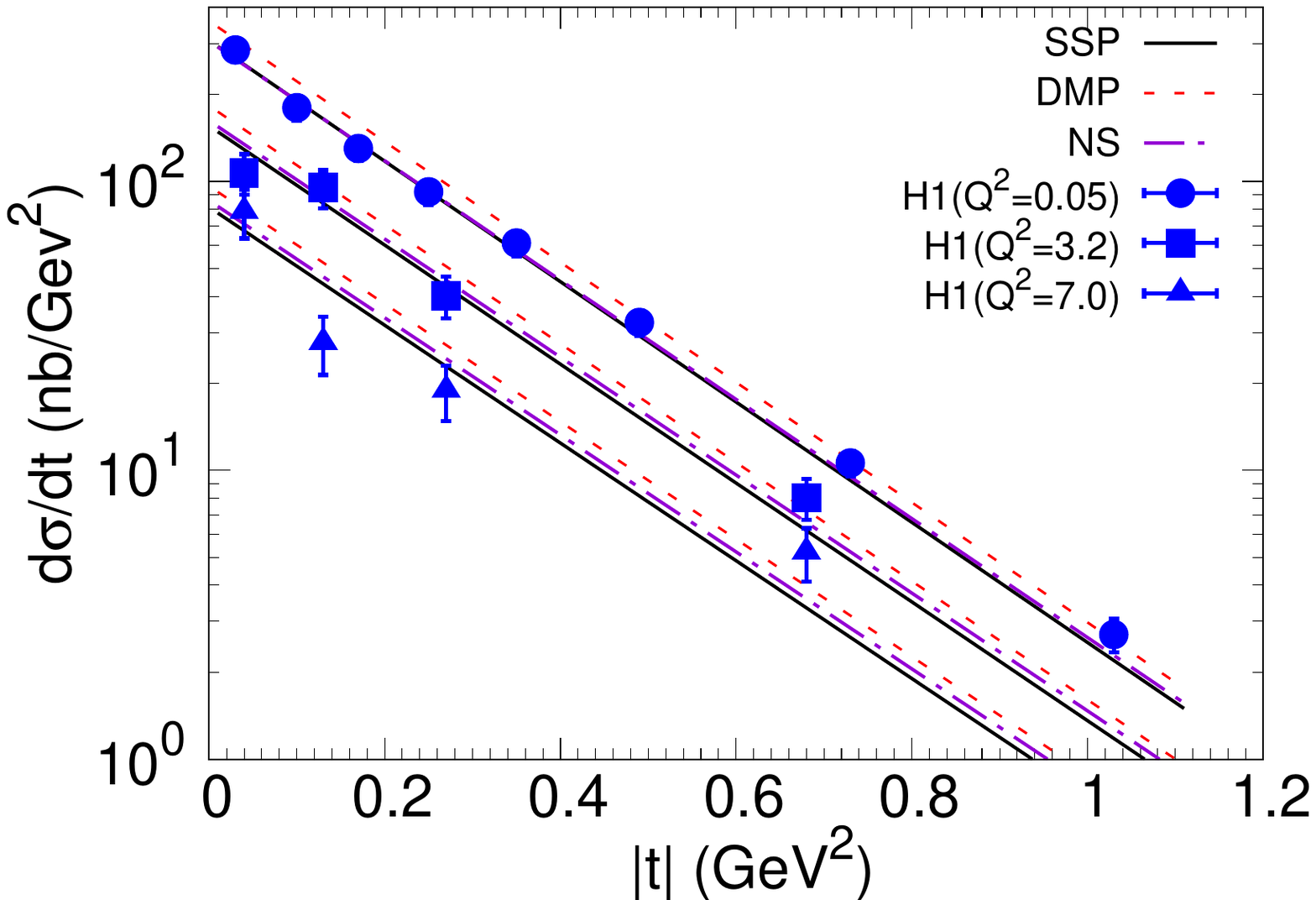}
\hspace{-0.3cm}
\includegraphics[width=0.48\textwidth]{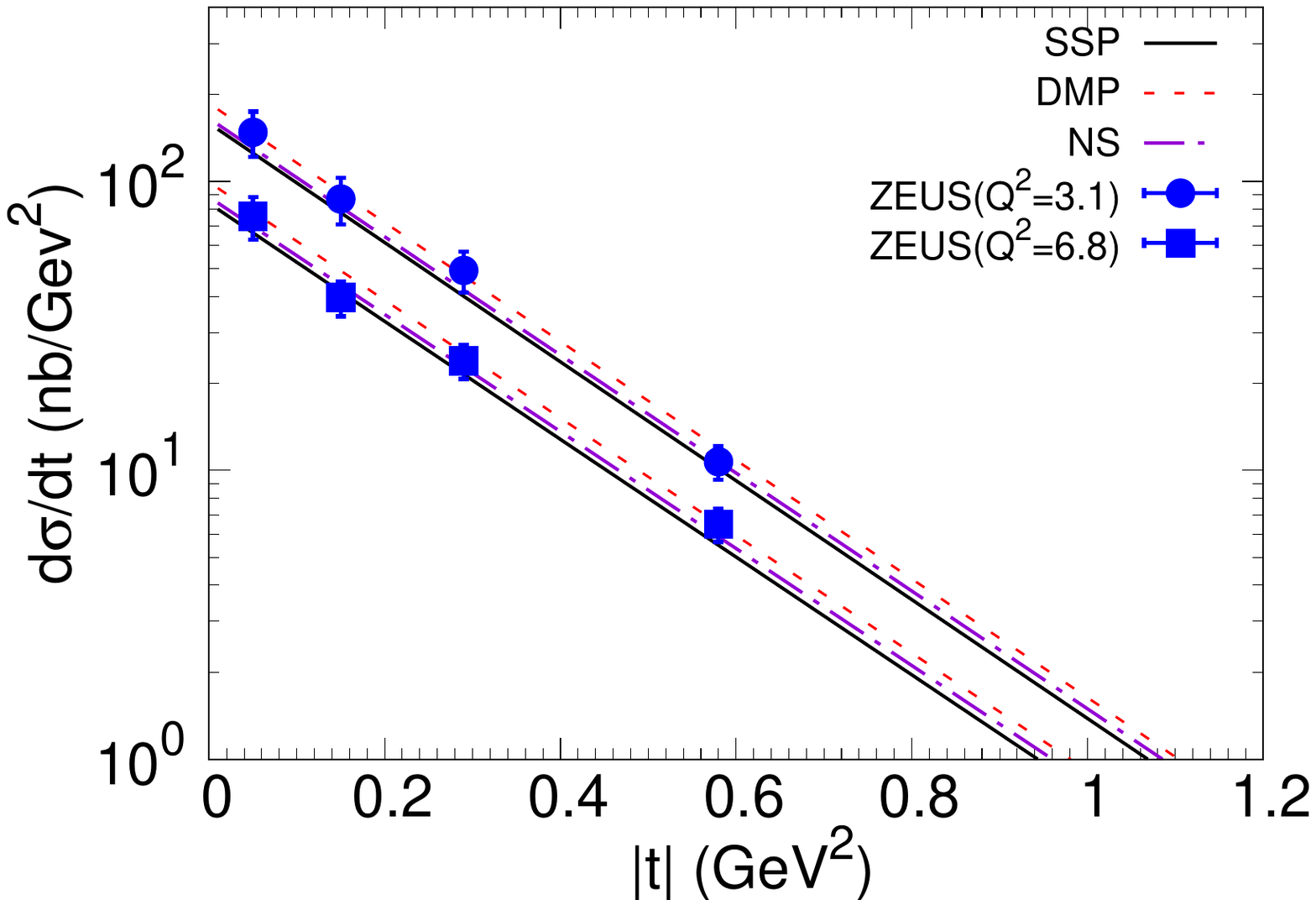}
\end{center}
\vspace{-0.5cm}
\caption{The differential cross section $d\sigma/dt$ for $J/\psi$ as a function of $|t|$ at different $Q^{2}$. The solid lines, the dashed lines, and the dash-dotted lines represent the results calculated from saturation scale prescription (SSP), the dipole momentum prescription (DMP), and the numerical results (NS), respectively. The data are from H1 \cite{H1:2005dtp} and ZUES \cite{ZEUS:2004yeh} Collaborations at HERA.}
\label{fig:t_dis}
\end{figure}

Figure \ref{fig:t_dis} shows the differential cross section for $J/\psi$ as a function of squared momentum transfer at different virtualities. The $J/\psi$ experimental data are taken from H1 \cite{H1:2005dtp} and ZUES \cite{ZEUS:2004yeh} Collaborations. In the simulation, the free parameters from the MPS model are $B=2.2\,\text{GeV}^{-2}$ and $\sigma_0= 68 \,\text{GeV}^2$. The dash-dotted lines (NS) donate the results from the numerical solution of the running coupling BK equation in previous study (similarly hereinafter) \cite{Cai:2020exu}. From Fig.\ref{fig:t_dis}, one can see that both presentations give a good description for the differential cross section expect for the data at $Q^{2}=7.0\,\text{GeV}^2$ with large uncertainty.

\begin{figure}[h!]
\begin{center}
\includegraphics[width=0.48\textwidth]{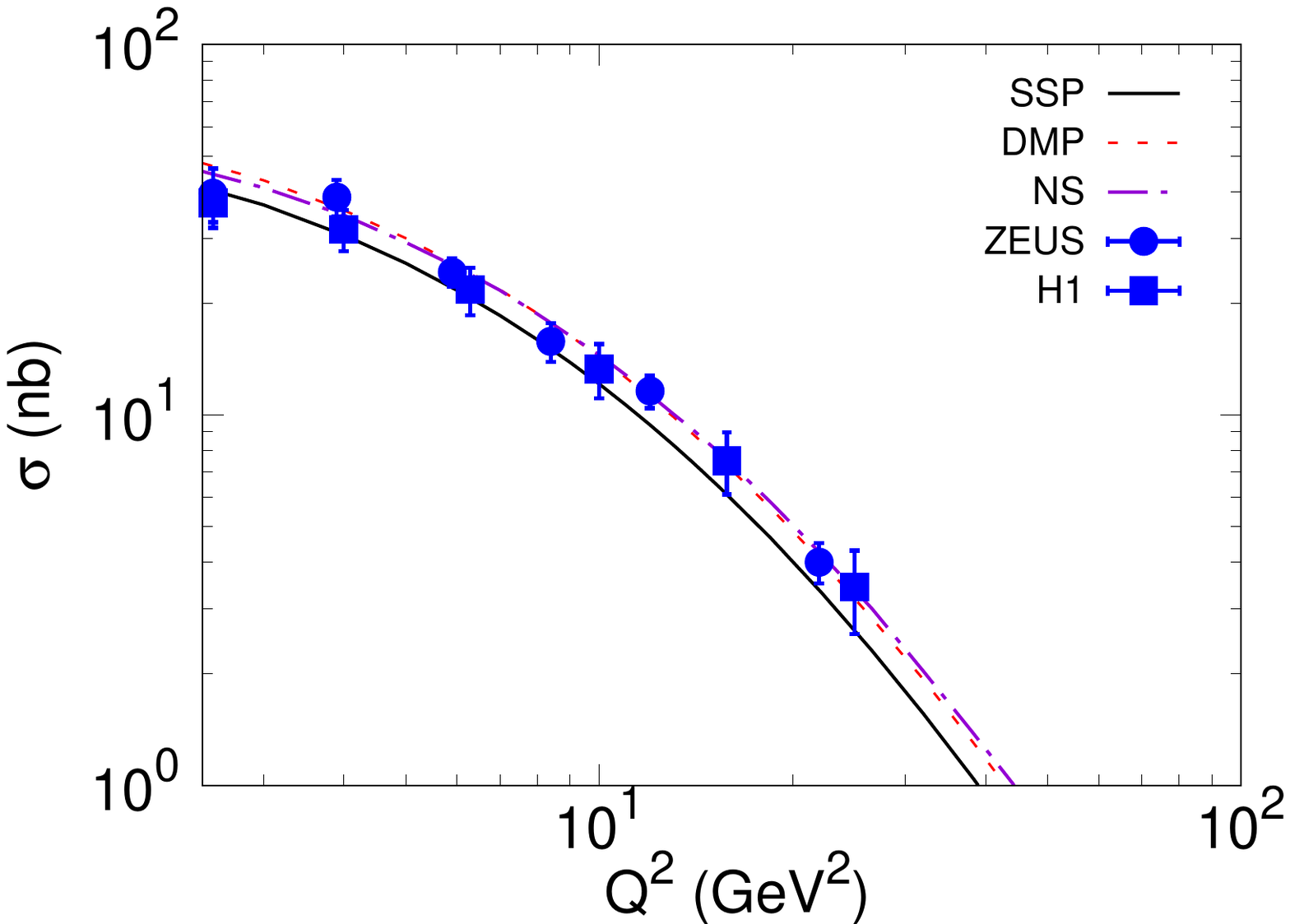}
\end{center}
\vspace{-0.5cm}
\caption{The total cross section $\sigma$ for $J/\psi$ as a function of $Q^{2}$. The solid lines, the dashed lines, and the dash-dotted lines represent the results calculated from saturation scale prescription (SSP), the dipole momentum prescription (DMP), and the numerical results (NS), respectively. The data are from H1 \cite{H1:2005dtp} and ZUES \cite{ZEUS:2004yeh} Collaborations at HERA.}
\label{fig:Q2_dis}
\end{figure}

\begin{figure}[h!]
\begin{center}
\includegraphics[width=0.48\textwidth]{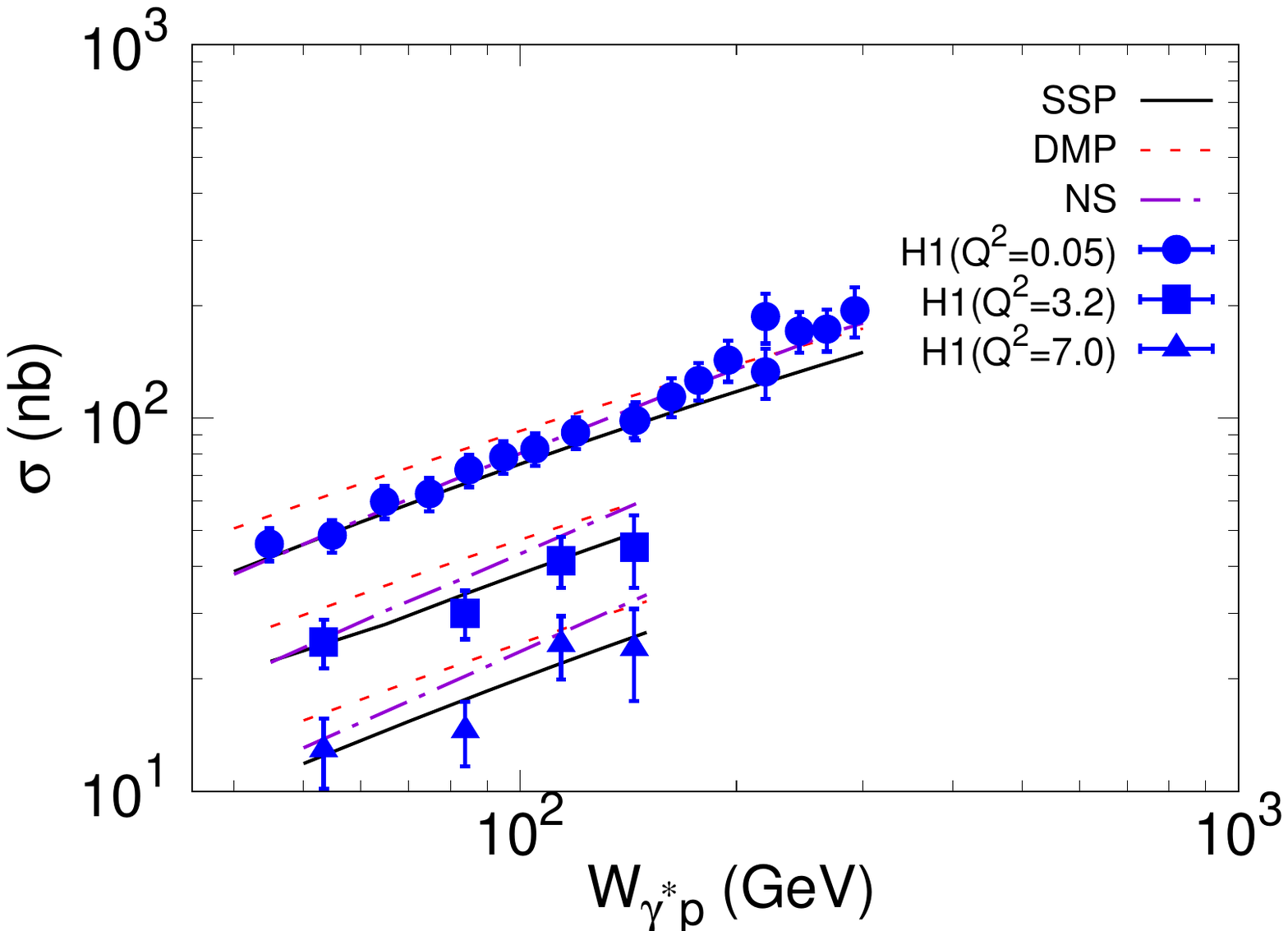}
\hspace{-0.3cm}
\includegraphics[width=0.48\textwidth]{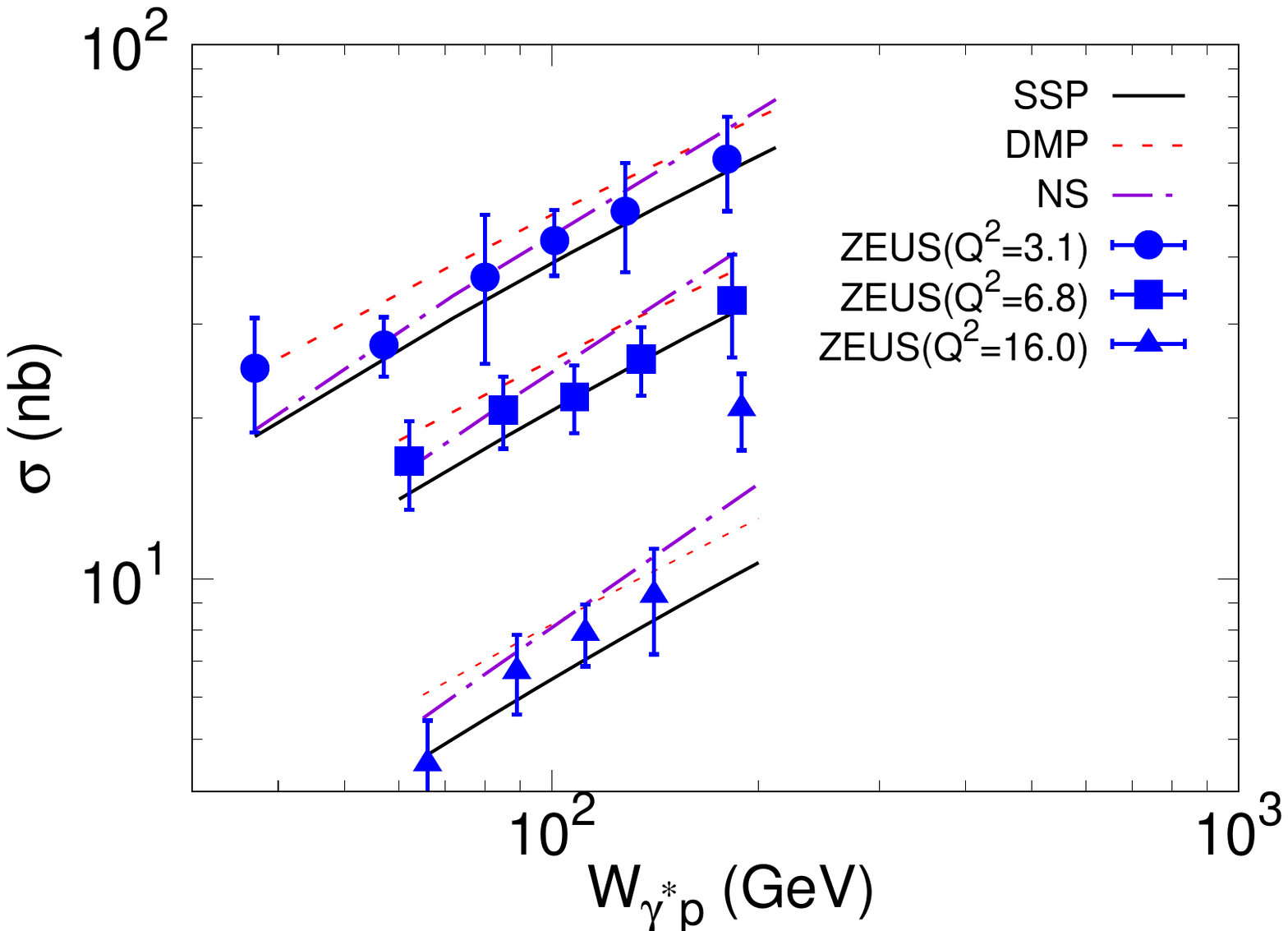}
\end{center}
\vspace{-0.5cm}
\caption{The total cross section $\sigma$ for $J/\psi$  as a function of $W_\gamma p$ at different $Q^{2}$. The solid lines, the dashed lines, and the dash-dotted lines represent the results calculated from saturation scale prescription (SSP), the dipole momentum prescription (DMP), and the numerical results (NS), respectively. The data are from H1 \cite{H1:2005dtp} and ZUES \cite{ZEUS:2004yeh} Collaborations at HERA.}
\label{fig:W_dis}
\end{figure}

We also present the total cross section for exclusive $J/\psi$ production. Figures.\ref{fig:Q2_dis} and Fig.\ref{fig:W_dis} present the predictions for the total cross section for $J/\psi$ as a function of virtuality and photon-hadron center-of-mass energy. The results from both presentations are in good agreement with the experimental data. As shown in Figs.\ref{fig:t_dis}-\ref{fig:W_dis}, the results obtained from the analytical solutions are consistent with those obtained from the numerical solutions. Note that the parameters from the dipole scattering amplitude are the same as the proton structure function. Therefore, our analytical solutions are robustness. These results indicate that our analytical solution of the BK equation with running constant can give a good description for the proton structure function and the exclusive vector meson production.

\section{conclusions and discussions}
An analytical solution for the BK equation, incorporating running coupling constant, has been derived utilizing the homogeneous balance method. The saturation scale dependent running coupling and dipole momentum dependent running coupling, which correspond to two distinct nonlinear evolution equations, have been considered. By applying the homogeneous balance principle, a heuristic solution with undetermined parameters is obtained. The experimental data of the proton structure function has been fitted to obtain the corresponding definitive solution for the BK equation, in different running coupling scale prescriptions. The $\chi^{2}/d.o.f$ values for saturation scale prescription and dipole momentum prescription are $1.07$ and $1.43$, respectively, indicating a scale dependent solution. However, the values have little difference in certain regions, suggesting that both prescriptions provide a good description for the proton structure function. The definitive solutions have been utilized to calculate exclusive $J/\psi$ production, which demonstrates that the analytical solutions with running coupling constant are in line with the $J/\psi$ differential and total cross section. These outcomes indicate the usefulness of the homogeneous balance method in investigating the solution of the nonlinear BK equation. Furthermore, the analytical solution for the BK equation with higher order corrections is worth exploring, as more accurate data will be collected from future Electron Ion Collider (EIC) \cite{Accardi:2012qut}, Large Hadron Electron Collider (LHeC) \cite{LHeC:2020van}, and Electron-ion Collider in China (EicC) \cite{Anderle:2021wcy,Chen:2020ijn}, facilitating the study of gluon saturation. Overall, this approach holds promise for providing a robust analytical solution to predict proton structure function and the exclusive vector meson production.

\begin{acknowledgments}
This work is supported by the Strategic Priority Research Program of Chinese Academy of Sciences under the Grant No. XDB34030301; the National Natural Science Foundation of China under Grant No. 12165004; the Guizhou Provincial Basic Research Program (Natural Science) under Grant No. QKHJC-ZK[2023]YB027; the Education Department of Guizhou Province under Grant No. QJJ[2022]016.
\end{acknowledgments}

\bibliographystyle{JHEP-2modlong}
\bibliography{refs}

\end{document}